\documentclass[
preprint,
superscriptaddress,
nofootinbib,
natbib,
amsmath,amssymb,
aps,
prb,
]{revtex4-1}

\usepackage{graphicx}
\usepackage{float} 
\usepackage[caption=false,font=footnotesize]{subfig}  
\usepackage{dcolumn}
\usepackage{bm}
\usepackage{verbatim}   
\usepackage{color}      
\usepackage{siunitx} 
\usepackage[capitalise]{cleveref} 
\usepackage{verbatim} 

\begin{document}

\title{First-principles study of the terahertz third-order nonlinear \\
response of metallic armchair graphene nanoribbons}

\author{Yichao Wang}
\affiliation{Electrical and Computer Engineering}
\author{David R. Andersen}
\affiliation{Electrical and Computer Engineering}
\affiliation{Physics and Astronomy\\The University of Iowa, Iowa City, IA 52242 USA}

\date{May 25, 2016}

\begin{abstract}
We compute the terahertz third-order nonlinear conductance of  metallic armchair
graphene nanoribbons using time-dependent perturbation theory. Significant enhancement of the 
intrinsic third-order conductance over the result for instrinsic 2D
single-layer graphene is observed over a wide range of temperatures.
We also investigate the nonlinear response of extrinsic metallic
acGNR with $|E_F| \ll \SI{200}{meV}$. We find that the third-order conductance
exhibits a strong Fermi level dependence at low temperatures. A third-order
critical field strength of between $\sim 1$ and $\SI{5}{kV/m}$ is computed for
the Kerr conductance as a function of temperature. For the
third-harmonic conductance, the minimum critical field is computed to be $\sim \SI{5}{kV/m}$.
\end{abstract}

\maketitle


\section{Introduction}
Graphene, a monolayer of carbon atoms arranged in a 2D honeycomb lattice, has
excellent electronic, mechanical, thermal and optoelectronic
properties.\cite{grreview}  The spectrum of graphene is described by the
massless Dirac equation. Due to the many unique properties of graphene, it is
considered a promising material for electronic device applications.\par

In the terahertz (THz) to far-infrared (FIR) spectral regime, the optical conductance of graphene
based systems has attracted much interest due to the ongoing search for viable THz
devices. Graphene is traditionally a poor conductor in the THz to FIR spectrum,
with universal conductivity $\sigma_0=e^2/(4 \hbar)$ leading to an
absorption of only 2.3\% at normal incidence per graphene
layer.\cite{nair08fine} However, graphene has a number of features that make it
an attractive nonlinear system to
study.\cite{nlr07,coherent10,4wavegr12,review2014} These include a tunable Fermi
level, and more importantly a linear dispersion relation near the Dirac
point.\cite{neto09,sarma11} This linear dispersion and the accompanying constant
Fermi velocity $v_F$ have led to the theoretical prediction of the generation of
higher-order
harmonics in graphene.\cite{nlr07} Mikhailov and Ziegler have developed a quasi
classical kinetic theory and a quantum theory on the third order nonlinear process in graphene.\cite{mikhailov2008nonlinear,mikhailov2012theory} Wright \textit{et.al.} \cite{wright09} adopted 
a time dependent perturbation theory to analyse the linear and third order nonlinear
 optical response of intrinsic 2D single layer graphene (2D SLG) with an applied electric field of 
approximately \SI{100}{kV/m}, which indicates that the strong nonlinear conductance makes graphene 
a potential candidate for THz photonic and optoelectronic devices. 
Ang \textit{et.al.} \cite{blg10,subgap11} investigated the nonlinear optical
conductivity of bilayer graphene (BLG), semihydrogenated graphene (SHG) and
Kronig-Penney (KP) graphene superlattices.  Gullans \textit{et.al.}
\cite{gullans2013single} studied the single photon nonlinear mechanism in
graphene nanostructures and showed strong confinement of plasmons and large
intrinsic nonlinearity in graphene nanostructures led to significant electric
field enhancement. Recently, Mikhailov \textit{et.al.}
\cite{mik14,*mik14R,mik16}, Cheng \textit{et.al.}
\cite{cheng1,*cheng1R,cheng2,cheng3,*cheng3R} and Morimoto \textit{et.al.}
\cite{scaling} proposed  quantum theories of the third-order nonlinear response
with an uniform external electric field in 2D SLG
independently. This work
\cite{mik14,*mik14R,mik16,cheng1,*cheng1R,cheng2,cheng3,*cheng3R,scaling}
studies the relationship of the Fermi energy with the direct interband
transition, which confirms the resonant frequencies for the third-harmonic
conductance which appeared in Refs. \cite{wright09,ang2015nonlinear}, and the
missing resonant frequencies for the Kerr conductance in Refs. \cite{wright09,ang2015nonlinear} as we perform the calculations of Refs. \cite{wright09,ang2015nonlinear}.\par

Hendry \textit{et.al.}\cite{coherent10} first report
measurement of the coherent nonlinear optical response of single and few-layer graphene 
using four-wave mixing.
Their results experimentally demonstrate that graphene structures exhibits
 a strong nonlinear optical response in the NIR spectral region.  Harmonic
 generation, frequency mixing, optical rectification, linear and circular photogalvanic effect, 
photon drag effect, photoconductivity, coherently controlled ballistic charge currents, etc. in
graphene are currently the subject of intense research, and have already found a number of
applications.\cite{review2014} Kumar \textit{et.al.} \cite{kumar2013third} found third harmonic generation in graphene and multi-layer graphite films grown by exfoliation. They found the nonlinear emission frequency matched well with the theoretical prediction and deducted an effective third order susceptibility on the order of \SI{100}{\micro m^2/kV^2}.  Maeng  \textit{et.al.} \cite{maeng2012gate} measured the nonlinear conductivity of gate controlled graphene grown by CVD. Their work show nonlinear conductance of graphene can be efficient controlled via applied gate voltage and doping. Recently, Hafez \textit{et.al.} \cite{noepgr15} reported experimental
results on the carrier dynamics in epitaxially grown monolayer graphene \cite{noepgr15}.  This work demonstrates that the microscopic mechanisms 
of nonlinear effects in graphene can be quite different from their counterparts in ordinary 
semiconductor systems\cite{noepgr15}. 
The large nonlinear response 
originating from interband transitions is seven orders of magnitude stronger than 
the nonlinear response observed in dielectric materials without such transitions\cite{coherent10,bao12}.
These theoretical and experimental studies have shown that
the linear energy dispersion and high electron Fermi velocity in graphene leads to a strongly 
nonlinear optical response in the THz to FIR regime for various 2D graphene systems 
compared with the counterparts in conventional parabolic semiconductor systems.\par

While the nonlinear optical properties of 2D graphene structures have been studied extensively,
the nonlinear optical response, which is proportional to the higher powers of the applied electric field, 
has been much less studied for graphene nanoribbons (GNR).  Duan \textit{et.al.}
\cite{duan2010infrared} studied the linear response of intrinsic metallic
armchair GNR in the infrared regime with a linearly-polarized applied
electric at low temperatures. Sasaki \textit{et.al.} \cite{gnrselectionrule}
proposed optical interband transition selection rules for acGNR with
linearly-polarized electric fields in the transverse and longitudinal
directions.  Chung \textit{et.al.}\cite{gnrselectionrule2} also investigated the
interband selection rules for acGNR. All of this work focused on the linear
response of GNR and did not address the nonlinear response of acGNR at THz
frequencies for an applied linearly-polarized electric field in the longitudinal
and transverse directions. \par

Wang \textit{et.al.}  \cite{GNRisQW} find that thin GNRs (sub-\SI{20}{nm}) with smooth edges can be treated as quasi 1D quantum wires, not dominated by defects. In general, new physics (quantization of energy,
momentum \textit{etc.}) emerges when the dimensionality of 2D graphene is
reduced to a quasi 1D quantum wire. With the rapid development of techniques for the synthesis of thin GNRs 
 \cite{GNRisQW,kimouche2015ultra,jacobberger2015direct}, thin GNRs (sub-\SI{20}{nm}) 
may have ultra smooth edges, higher mobility and longer carrier mean free path
than expected theoretically.  Depending on the nature of the edges,
there are two types of GNR: armchair graphene nanoribbons (acGNR) and the zigzag graphene
nanoribbon (zzGNR). Electron dynamics of both acGNR and zzGNR have distinct
properties, due to their geometry and boundary conditions.\cite{fertig1,
fertig2}
Metallic acGNR exhibits a linear band structure in both tight-binding
\cite{and1,and2} and $\mathbf{k} \cdot \mathbf{p}$ models.
Edge states contribute significantly to GNR properties, since in a nanoscale GNR, 
massless Dirac fermions can reach the ribbon edge within a few femtoseconds before encountering 
any other scattering and screening effects, such as electron-electron and electron-phonon
interactions, the Peierls instability, \textit{etc.}  In general, the nonlinearity of GNR originates 
from the redistribution of the Dirac fermions in momentum and energy space induced 
by the applied electric field\cite{review2014}. As a consequence, conductivity components oscillating 
in time and space, as well as spatially homogeneous steady state components are expected to be 
obtained from the resulting nonequilibrium distribution. Thus the resulting nonlinear response is sensitive to the applied field strength and polarization\cite{review2014}. Therefore, it is important to 
study the electrodynamics for higher order harmonic generation with the existence
of an applied electric field in GNR. In light of recent reports of the growth of ultra thin acGNR (sub-\SI{10}{nm}) reported by Kimouche \textit{et.al.} \cite{kimouche2015ultra} and Jacobberger \textit{et.al.} \cite{jacobberger2015direct}, and
the fact that Kimouche \textit{et.al.}\cite{kimouche2015ultra} show that defects
(kinks) do not strongly modify the electronic structure of ultrathin acGNR, the study of
the nonlinear response of these metallic acGNR is of particular significance
today.

In this paper, we develop a semi-analytic approach based on the $\mathbf{k}
\cdot \mathbf{p}$ approximation in the Coulomb gauge to calculate the nonlinear
THz response of thin acGNR (width $<\SI{20}{nm}$) under a moderate applied linearly-polarized electric field
in the longitudinal  and transverse directions. We use time dependent
perturbation theory to do a Fourier analysis of the wavefunction in the presence
of a strong linearly-polarized time-harmonic electric field, and obtain the linear and third-order optical THz response of thin metallic acGNR. \par

The paper is organized as follows. In Sec. \ref{Model}, we begin with the
$\mathbf{k} \cdot \mathbf{p}$ approximation to obtain the time-independent wave
equation and the interaction Hamiltonian with an applied electric field for
acGNR, and we present a brief derivation of our semi-analytical approach to
calculate the nonlinear conductance.  In Sec. \ref{Results}, we apply our model to calculate the
nonlinear conductance of metallic acGNR. In particular, we compare the
nonlinear properties of single layer metallic acGNR with those of intrinsic 2D
SLG. We also propose a correction to previous
work\cite{wright09,ang2015nonlinear} on the third order Kerr conductance in intrinsic 2D SLG. We analyze the
third-order nonlinear terms using standard definitions for these quantities:
Kerr conductance for the third-order terms oscillating at frequency $\omega$ and
third-harmonic conductance for the terms oscillating at frequency $3 \omega$, determine the required applied electric
field strength to induce non-negligible nonlinear effects and investigate
the temperature and Fermi level dependence of the
nonlinear conductance. Following this, a brief analysis of the selection rules
for nonlinear \si{THz} direct interband transitions in metallic thin acGNR is discussed. Finally, the conclusions are
presented in Section \ref{Conclusions}.


\section{Model}\label{Model}
\subsection{$H_0$, $\psi_0$ and the applied field $E_{\mu}$}
Graphene is a 2D hexagonal lattice (honeycomb) structure of covalently bonded
carbon atoms.  As there are 2 atoms per unit cell, we label them A and B
respectively. At low energies, graphene carriers can be described by the
massless Dirac equation. As a consequence, graphene shows a linear energy band
structure near the Dirac points $\mathbf{K}=\frac{2 \pi}{a_0}\left ( \frac{1}{3} ,\frac{1}{{\sqrt{3}}}\right )$ and $\mathbf{K'}=\frac{2 \pi}{a_0}\left ( \textnormal{-}\frac{1}{3} ,\frac{1}{\sqrt{3}}\right )$ of the Brillouin zone. Here $a_0$ is the triangular lattice parameter of the graphene structure.\cite{fertig1,fertig2}
($a_0 = \sqrt{3} a_{cc}$ where $a_{cc}$ is the carbon-carbon separation distance
in acGNR and $a_{cc} = 1.42 \, \mathrm{\AA}$).\par

The unperturbed $\mathbf{k} \cdot \mathbf{p}$ Hamiltonian for graphene 
can be written in terms of Pauli matrices as
$H_{0,K} = \hbar v_F \bm{\sigma} \cdot \mathbf{k}$ for the $\mathbf{K}$ valley
and $H_{0,K'} = \hbar v_F \bm{\sigma} \cdot \mathbf{k'}$ for the $\mathbf{K'}$
valley with $\mathbf{k}(\mathbf{k'})$ the perturbation from the center of
the $\mathbf{K}(\mathbf{K'})$ valley. The corresponding wave functions are expressed as envelope functions $\psi_{K}(\mathbf{r})=\left [ \psi _A(\mathbf{r}), \psi _B(\mathbf{r})\right ]$ and $\psi_{K'}(\mathbf{r})=\left [ \psi _A' (\mathbf{r}), \psi _B' (\mathbf{r})\right ]$ for states near the $\mathbf{K}$ and $\mathbf{K'}$ points, respectively. \par

Following the development in \cite{fertig1,fertig2}, the time-independent (unperturbed)
Hamiltonian for a single Dirac fermion in GNR can be written as:
\begin{equation}
\begin{aligned}
H_0&=\begin{pmatrix} H_{0,K}&0\\0&H_{0,K'}\end{pmatrix}\\
&=\hbar  v_F
\begin{pmatrix}
 0 & k_x-i k_y & 0 & 0 \\
 k_x+i k_y & 0 & 0 & 0 \\
 0 & 0 & 0 & -k_x-i k_y \\
 0 & 0 &-k_x+i k_y & 0  \end{pmatrix}
\end{aligned}\label{eq:H0}
\end{equation}
with wave envelope functions in the case of acGNR:
\begin{equation}\label{eq:psi0}\psi_{n,s}(\mathbf{r},0)=\begin{pmatrix} \psi_{n,s}(\mathbf{r})_K\\\psi_{n,s}(\mathbf{r})_{K'}\end{pmatrix}=\frac{e^{i k_y y}}{2\sqrt{L_x L_y}}
\begin{pmatrix} e^{\textnormal{-} i \theta_{k_n,k_y}} e^{i k_n x}\\s e^{i k_n
x}\\-e^{\textnormal{-} i \theta_{k_n,k_y}} e^{\textnormal{-}i k_n x}\\s
e^{\textnormal{-}i k_n x}\end{pmatrix}
\end{equation}
with $L_x$ the width of acGNR in the $x$ (zigzag) direction,  $L_y$ the length of
the acGNR in
the $y$ (armchair) direction and the direction of the isospin of the state is $\theta_{k_n,k_y}=\tan^{-1} (k_n/k_y)$.

\begin{figure}[H]
\centering
\includegraphics[width=1.0\textwidth]{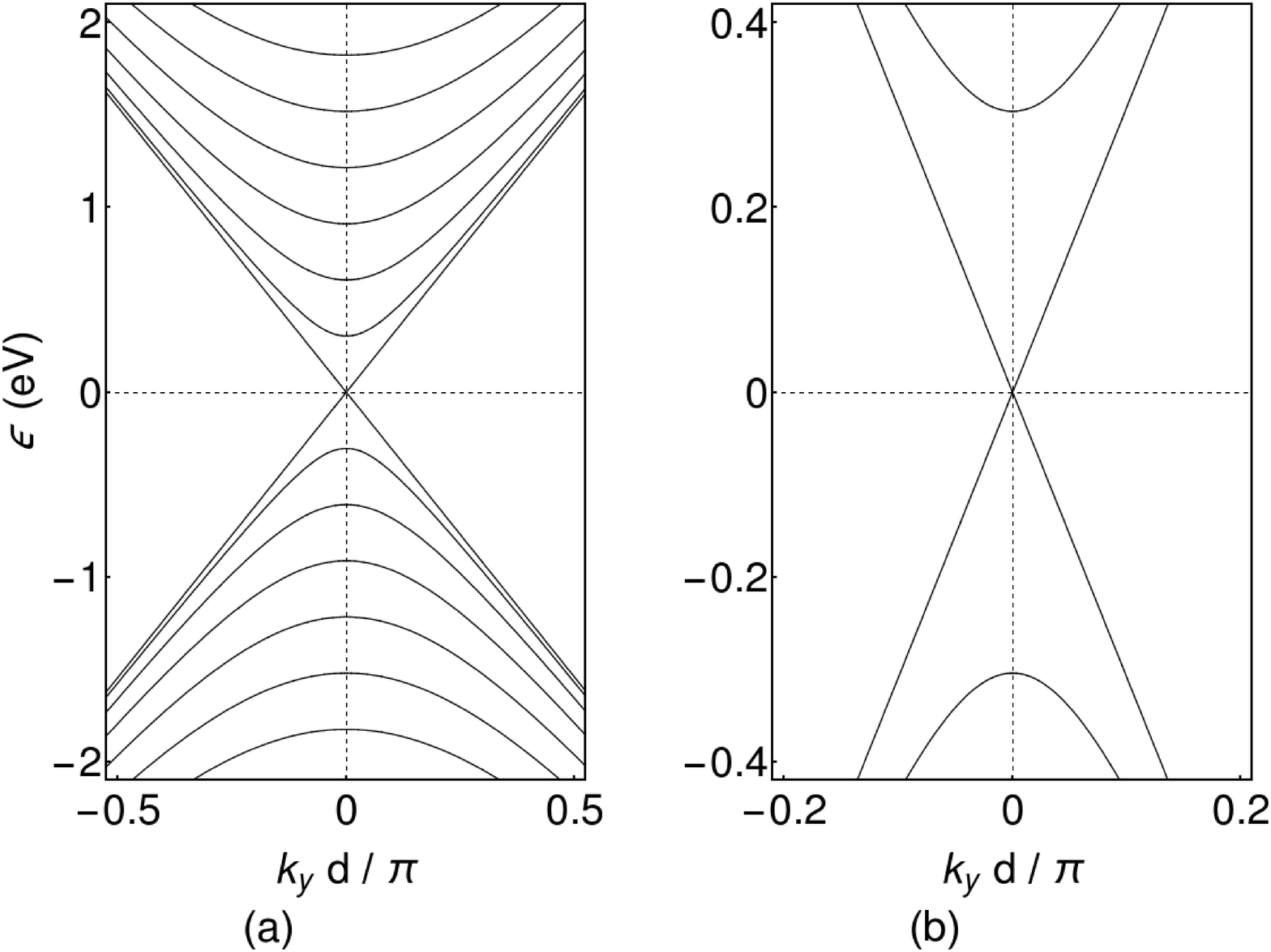}
\caption{$\mathbf{k} \cdot \mathbf{p}$ band
structure of infinitely long metallic acGNR of width $L_x = 24.6 \mathrm{\AA}$
(acGNR20) and $L_y \to \infty$.
(a) illustrates the seven lowest-energy bands, and (b) illustrates the gap of
$\sim 608\, \mathrm{meV}$ between $n=1$ conduction and valence band.
Here $d$ is the width of the acGNR unit cell ($d = (1 + \sqrt{3}) a_{cc}$).
}
\label{fig:1}
\end{figure}

The electronic properties of acGNR depend strongly on their width $L_x$. The
width of acGNR can be calculated using $L_x=\frac{N}{2} a_0$, where $N$ is the number of atoms along the zigzag edge ($\hat{x}$ direction). In general, acGNR of $N=3M-1$ atoms wide along the zigzag edge, with \textit{M} odd, are
metallic, whereas all the other cases are semiconductors \cite{fertig1, fertig2}. In Fig.
\ref{fig:1} we plot the band structure of infinitely long metallic $(L_y \to
\infty)$ acGNR for $N=20$ (acGNR20). One can
see that in Fig. \ref{fig:1} there is a Dirac point, leading to metallic
behavior for a single-electron model. Thus for a width of the form
$L_x=\frac{3M-1}{2} a_0$ with $M$ odd, the allowed values of $k_n=\frac{2\pi}{3
a_0}\frac{M+n}{M}$ create doubly-degenerate states for $n\neq -M$
and when $k_y \to 0$, the existence of a zero energy state indicates
that the conduction and valence band touch at the Dirac points.
The non-metallic bands in \ref{fig:1} are well above THz energies, and as a
result, a THz direct interband transition can only occur between metallic
subbands ($k_n = 0$) for thin metallic acGNR.

Because thin acGNR (sub-$\SI{20}{nm}$) can be treated as a
quasi-1D quantum wire system \cite{GNRisQW}, we have Bloch states where
$k_{x,n}=\frac{2\pi}{3 a_0}\frac{M+n}{M}$ and $k_{y,m}=\frac{2\pi}{L_y}m$. In metallic acGNR when $n=-M$,
we can write the time-independent wave envelope function for one Dirac fermion
in the lowest subband near the Dirac point, with $k_{x,n}=0$ as:
\begin{equation}\label{eq:psir0m}
\psi(\mathbf{r},0;m)= \phi_0(m)e^{i 2\pi m y/L_y}
\end{equation}
where $\phi_0(m)$ is found to be:
\begin{equation}\label{i.c.}
\phi_0(m)=\begin{bmatrix}\phi_{K,0}(m)\\\phi_{K',0}(m)\end{bmatrix}=\frac{1}{2 \sqrt{L_x L_y}}\begin{bmatrix} \mathrm{sgn}(k_y)\\ s\\
-\mathrm{sgn}(k_y) \\s\end{bmatrix}\end{equation} 
constructed from Eq. \ref{eq:psi0}.\par

Let us consider metallic acGNR under an applied linearly-polarized electric field $\mathbf{E} = \hat{\mu}
E_\mu e^{-i \omega t}$, of frequency $\omega$ with normal incidence. Notice that the time dependent part of the
applied field $e^{-i \omega t}$ corresponds to the absorption process and $e^{i \omega
t}$ corresponds to the emission process. For time-harmonic fields that turn on
adiabatically \cite{mahan,ryu07} at $ t \to -\infty$ and constant scalar
potential $\nabla \varphi=0$, in the Coulomb
gauge \cite{mahan} ($\nabla \cdot \mathbf{A}=0$) the vector potential
\cite{mahan,wright09,ang2015nonlinear} is of the
form $\mathbf{A} = \hat{\mu} E_\mu \exp(-i \omega t)/(i \omega)$
(see Appendix \ref{app1} for a brief discussion).
The interaction with the vector potential is described by
writing the canonical momentum $\mathbf{k} \rightarrow \mathbf{k} +\frac{q
\mathbf{A}}{\hbar}$, where $q$ is the elementary charge.
In other words, the total Hamiltonian for graphene in the presence of a
normally-incident electromagnetic field
can be written as $H_K =\hbar v_F \bm{\sigma} \cdot (\mathbf{k} +\frac{q
\mathbf{A}}{\hbar})$ for the $\mathbf{K}$ point and  $H_{K'} =\hbar v_F \bm{\sigma}
\cdot (\mathbf{k'} +\frac{q \mathbf{A}}{\hbar})$ for the $\mathbf{K'}$ point. The total Hamiltonian for
acGNR
can be expressed as: $H=H_0+H_{int}$, where the interaction
part of the Hamiltonian is given by:
\begin{equation}
H_{int}=\begin{pmatrix} H_{int,K}&0\\0&H_{int,K'}\end{pmatrix}
\end{equation}
with $H_{int,K(K')}=\frac{q v_F}{i \omega} \bm{\sigma} \cdot \mathbf{E_0} e^{-i \omega t}$
where $\bm{\sigma}=\hat{x} \sigma_x +\hat{y} \sigma_y$ is the Pauli matrix and
$\mu = x, y$ indicates the direction of the applied linearly-polarized electric field.

\subsection{Local conductivity and conductance}
In this work, we follow Refs.
\cite{ryu07,eo10,wright09,blg10,subgap11,cheng1,*cheng1R,cheng2,cheng3,*cheng3R,mik16,ang2015nonlinear,scaling}
and make the relaxation-free approximation,
neglecting carrier-phonon and carrier-carrier \cite{tau13} scattering,
defect scattering,
and many body effects in our calculation.
Acoustic phonon scattering may be neglected because the interaction is not
phasematched due to the large (three orders of
magnitude) difference between the carrier Fermi velocity $v_F$ and the acoustic
velocity. The optical phonon energy in graphene is $\sim \SI{200}{meV}$ and so
for low-energy carriers of the order of a few tens of meV and below, optical
phonon scattering may be neglected as well.
Carrier-carrier scattering increases with the square of the carrier density.
Since our model considers extrinsic metallic acGNR with Fermi energies of the
order of a few meV and small excitation field strengths ($\sim \SI{10}{kV/m}$),
carrier-carrier scattering and many-body effects may be neglected to a good approximation.
Ultrathin metallic acGNR with smooth edges have recently been fabricated showing
ballistic transport due to the low defect density \cite{kimouche2015ultra}, and
so it is appropriate to neglect
defect scattering. Due to the block nature of
the total Hamiltonian $H=H_0+H_{int}$ in the $\mathbf{k} \cdot \mathbf{p}$ approximation,
we also neglect intravalley and intervalley scattering in thin metallic acGNRs as well.
Thus, the theory presented in this paper applies to low-energy (THz) carriers in
thin, smooth metallic acGNR where the higher index bands $(k_{x,n} \neq 0)$ are well-separated
from the lowest-order linear bands (see Fig. \ref{fig:1}).

In metallic acGNR, we describe the Dirac fermion under the influence of an
applied electric field $\hat{\mu} E_{\mu}e^{-i\omega t}$ for the metallic band ($k_{x,n}=0$) as an envelope wave function $\psi_{\mu}(\mathbf{r}, t;m)=\left [
\psi _{\mu}(\mathbf{r},t;m)_K, \psi _{\mu}(\mathbf{r},t;m\right)_{K'}]^T$.
Using the Floquet theorem, the Fourier series expansion of $\psi_{\mu}(\mathbf{r}, t;m)$
 can be written \cite{wright09,ang2015nonlinear,lopez08,blg10,subgap11,zhou11} as:
\begin{equation}\label{eq:psi}
\psi_{\mu}(\mathbf{r}, t;m)=\sum_{l=0}^{\infty} \phi_{\mu}(m,l) e^{i 2\pi m y/L_y} e^{\pm i \omega l t} e^{-i \epsilon t / \hbar}
\end{equation}
with the initial condition $\phi_{\mu}(m,0)=\phi_0(m)$, which satisfies the
requirement that when $\mathbf{A}\to 0$, $\psi_{\mu}(\mathbf{r}, t;m)$ should be
a solution of the Hamiltonian without an applied
field\cite{wright09,ang2015nonlinear,lopez08,blg10,subgap11}. The spinor
$\phi_{\mu}(m,l)$ is given by:
\begin{equation}
\phi_{\mu}(m,l)=\begin{bmatrix} \phi_{\mu}(m,l)_{K}\\ \phi_{\mu}(m,l)_{K'} \end{bmatrix}=\begin{bmatrix} a_{l}(m)\\b_{l}(m)\\c_{l}(m)\\d_{l}(m) \end{bmatrix}
\end{equation}

We can then calculate the
charge density as: $\mathbf{\rho}=|\psi_{\mu} (\mathbf{r},t;m)|^2$, where the particle density
operator is  $\rho_{op}(\mathbf{r})=\delta(\mathbf{r-r_{op}})$. 
After applying the continuity equation $q \frac{\partial \rho}{\partial
t}+\mathbf{\nabla \cdot j}=0$, along with the Schr\"{o}dinger equation $H
\psi_{\mu}(\mathbf{r}, t;m)=i \hbar \frac{\partial \psi_{\mu}(\mathbf{r},
t;m)}{\partial t}$ under the Coulomb gauge, we
obtain the local (single-particle) current density for Dirac fermions in the
metallic sub-band of acGNR:
\begin{equation}
\mathbf{j}(m,t)=\hat{x} j_{x}(m,t)+\hat{y} j_{y}(m,t)
\end{equation}
with the local current density component defined as:
\begin{equation}
j_{\nu}(m,t)=q \psi_{\mu}(\mathbf{r}, t;m)^\dagger
\frac{\partial H}{\hbar \partial  k_\nu} \psi_{\mu}(\mathbf{r}, t;m)
\label{eq:jk}
\end{equation}
where $\mu=x,y$ indicates the
direction of the polarization of the applied electric field,
 and $\nu=x,y$ indicates the component of the induced current.
After substituting \cref{eq:psi} into \cref{eq:jk}, the Fourier series expansion
of the local current density becomes:
\begin{equation}\label{eq:fourierexpansion}
\begin{aligned}
j_{\nu}(m,t)=&q \left[\phi_{\mu}(m,0)+\phi_{\mu}(m,1)e^{-i\omega t}+\phi_{\mu}(m,2)e^{-i2\omega t}+\cdots \right]^{\dagger} \\
&\times \frac{\partial H}{\hbar \partial  k_\nu} \left[\phi_{\mu}(m,0)+\phi_{\mu}(m,1)e^{-i\omega t}+\phi_{\mu}(m,2)e^{-i2\omega t}+\cdots \right] \\
=&q \left\{\left[\phi_{\mu}^{\dagger} (m,0)\frac{\partial H}{\hbar \partial
k_\nu}\phi_{\mu}(m,0)+\phi_{\mu}^{\dagger} (m,1)\frac{\partial H}{\hbar \partial
k_\nu}\phi_{\mu}(m,1)+\cdots\right]\right. \\
&+e^{-i \omega t} \left[\phi_{\mu}^{\dagger} (m,0)\frac{\partial H}{\hbar \partial  k_\nu}\phi_{\mu}(m,1)+\phi_{\mu}^{\dagger} (m,1)\frac{\partial H}{\hbar \partial  k_\nu}\phi_{\mu}(m,2)+\cdots\right] \\
&+e^{+i \omega t} \left[\phi_{\mu}^{\dagger} (m,1)\frac{\partial H}{\hbar \partial  k_\nu}\phi_{\mu}(m,0)+\phi_{\mu}^{\dagger} (m,2)\frac{\partial H}{\hbar \partial  k_\nu}\phi_{\mu}(m,1)+\cdots\right] \\
&+e^{-i 2\omega t} \left[\phi_{\mu}^{\dagger} (m,0)\frac{\partial H}{\hbar \partial  k_\nu}\phi_{\mu}(m,2)+\phi_{\mu}^{\dagger} (m,1)\frac{\partial H}{\hbar \partial  k_\nu}\phi_{\mu}(m,3)+\cdots\right]\\
&+e^{+i 2\omega t} \left[\phi_{\mu}^{\dagger} (m,2)\frac{\partial H}{\hbar \partial  k_\nu}\phi_{\mu}(m,0)+\phi_{\mu}^{\dagger} (m,3)\frac{\partial H}{\hbar \partial  k_\nu}\phi_{\mu}(m,1)+\cdots\right] \\
&+e^{-i 3\omega t} \left[\phi_{\mu}^{\dagger} (m,0)\frac{\partial H}{\hbar \partial  k_\nu}\phi_{\mu}(m,3)+\cdots\right]\\
&+\left . e^{+i 3\omega t} \left[\phi_{\mu}^{\dagger} (m,3)\frac{\partial
H}{\hbar \partial  k_\nu}\phi_{\mu}(m,0)+\cdots\right] +\cdots \right \}
\end{aligned}
\end{equation}

In general, for the study of third-order nonlinear optical processes induced by
an arbitrary superposition of three time-harmonic electric fields, it is
customary to write the local current density due to an individual atom as
the product of a fourth-rank conductivity tensor with the three arbitrary applied fields.
In the current work, we consider a much simpler case. The applied electric field
is linearly-polarized along the longitudinal armchair (transverse zigzag) or
$\hat{y}$ ($\hat{x}$) direction and has a single frequency $\omega$. As a
result, the expression for the local current density can be
written\cite{mik16}:
\begin{equation}\label{eq:localtensorexpansion}
\begin{aligned}
j_{\nu}(m,t)&=\left[e^{-i \omega t}
\tilde{\sigma}_{\mu\nu}^{(1)}(\omega)E_{\mu}+3e^{-i\omega t}
\tilde{\sigma}_{\nu\mu\mu\mu}^{(3)}(\omega,\omega,-\omega) E_{\mu}^3+e^{-i
3\omega t} \tilde{\sigma}_{\nu\mu\mu\mu}^{(3)}(\omega,\omega,\omega)
E_{\mu}^3+\cdots\right]+c.c. \\
&=\left[ j_\nu^{(1)}(m,\omega,t)+ j_\nu^{(3)}(m,\omega,t)
+ j_\nu^{(3)}(m,3 \omega,t )+ \ldots \right ] + c.c.
\end{aligned}
\end{equation}
By matching term-by-term the expansions in Eqs. \ref{eq:fourierexpansion} and
\ref{eq:localtensorexpansion}, we can obtain the individual non-zero elements in the
local third-order conductivity tensor. Further, by rewriting Eq.
\ref{eq:localtensorexpansion}, we see that the expressions for the Fourier
components of the local current density reduce to terms involving a local
$2\times2$ conductivity matrix and the applied electric
field\cite{wright09,ang2015nonlinear,blg10,subgap11}:
\begin{equation}
j_\nu^{(i)}(m,\omega_0) = \sigma_{\mu\nu}^{(i)}(m, \omega_0) E_\mu e^{-i
\omega_0
t} \label{eq:jm}
\end{equation}
where for $i=1$, $\omega_0 = \omega$; and for $i=3$, $\omega_0 = \omega$
($\omega_0 = 3 \omega$)
for the Kerr (third-harmonic) terms in the local current density
expansion, and
where $\sigma_{\mu\nu}^{(i)}(m,\omega_0)$ is the local \textit{i}th-order conductivity matrix
defined as for 2D SLG in Refs.
\cite{wright09,ang2015nonlinear,lopez08,blg10,subgap11,lopez08,mishchenko09,bao12,zhou11}

To compute the \textit{total} current density, we sum over all possible states,
using the thermal distribution
$N(\epsilon, E_F)=n_F(-|\epsilon|, E_F)-n_F(|\epsilon|, E_F)$ where $|\epsilon|=|m|hv_F/L_y$.
The total current density
\cite{wright09,ang2015nonlinear,lopez08,blg10,subgap11,lopez08,mishchenko09,bao12,zhou11}
is therefore:
\begin{equation}
J_{\nu}(t)=g_s \, g_v \sum_{m} j_{\nu}(m,t) N(\epsilon, E_F)
\end{equation}
with $g_s, \, g_v=2$ the spin and valley degeneracies respectively.
Here the initial occupancy of the system is described by the Fermi function
$n_F(\epsilon, E_F)$. Conduction band states are occupied with probability
$n_F(|\epsilon|, E_F)$ and valence band states are occupied with probability
$n_F(-|\epsilon|, E_F)$.
The Brey-Fertig wavefunction of \cref{eq:psir0m,i.c.} is normalized over the
entire sample\cite{fertig1,fertig2}, implying that
the states at $k_y$ for each valley are occupied with probability $1/2$ (assumes
$N$ carriers per unit cell). Since
there are 2$N$ carriers per unit cell, we multiply by $g_v = 2$ to include the
contribution to the total current from all 2$N$ carriers.
As the local current density
$j_{\nu}(m,t)$ conserves charge current density\cite{jk65,jk06,ryu07}  with an
applied vector potential $\mathbf{A}$ and the symmetry of graphene, it is straightforward to expand the total current component $J_{\nu}(t)$ as Fourier series of odd higher-harmonics \cite{wright09,ang2015nonlinear,lopez08,blg10,subgap11,nlr07,mikhailov2008nonlinear,coherent10,mikhailov2012theory,mik14,*mik14R,mik16,cheng1,*cheng1R,cheng2,cheng3,*cheng3R}.
Again, following Refs. \cite{mik16}, we write the total current density as:
\begin{equation}
\begin{aligned}
J_{\nu}(t)&=\left[e^{-i \omega t} \sigma_{\mu\nu}^{(1)}(\omega)E_{\mu}+3e^{-i\omega t} \sigma_{\nu\mu\mu\mu}^{(3)}(\omega,\omega,-\omega) E_{\mu}^3+e^{-i 3\omega t} \sigma_{\nu\mu\mu\mu}^{(3)}(\omega,\omega,\omega) E_{\mu}^3+\cdots\right]+c.c. \\
&=\left[J_{\nu}^{(1)}(\omega,t)+J_{\nu}^{(3)}(\omega,t)+J_{\nu}^{(3)}(3\omega,t)+\cdots\right]+c.c
\end{aligned}
\end{equation}
Adopting the notation in Refs.
\cite{wright09,ang2015nonlinear,blg10,subgap11,gnrselectionrule}, we define the
\textit{i}th-order conductance
component\cite{wright09,ang2015nonlinear,blg10,subgap11}
as a $2\times2$ conductance matrix relating the total nonlinear current density
and the applied linearly-polarized electric field:
\begin{equation}
J_{\nu}^{(i)}(\omega_0,t)=g_{\mu\nu}^{(i)}(\omega_0)E_{\mu} e^{-i \omega_0 t}
\end{equation} 

For the metallic band in thin acGNR, with an applied a $\hat{y}$-polarized
electric field $\hat{y} E_y e^{-i \omega t}$, the Hamiltonian $H$ for $k_y=2\pi m/L_y$ can be written as:
\begin{equation}
\begin{aligned}
H & =H_0+H_{int}\\
&=\hbar  v_F
\begin{pmatrix}
 0 & -i (k_y+\frac{e E_y}{i \hbar \omega} e^{-i \omega t}) & 0 & 0 \\
 +i (k_y+\frac{e E_y}{i \hbar \omega} e^{-i \omega t}) & 0 & 0 & 0 \\
 0 & 0 & 0 & -i (k_y+\frac{e E_y}{i \hbar \omega} e^{-i \omega t}) \\
 0 & 0 &+i (k_y+\frac{e E_y}{i \hbar \omega}e^{-i \omega t}) & 0  \end{pmatrix}
\end{aligned}\label{eq:Hfinal}
\end{equation}
We then proceed to solve the Schr\"{o}dinger equation $H
\psi_{\mu}(\mathbf{r}, t;m)=i \hbar \frac{\partial }{\partial
t}\psi_{\mu}(\mathbf{r}, t;m)$. Due to the orthogonal properties of the basis sets
\{$e^{-i l \omega t}$\}, we obtain the following recursion relations for the spinor
components:
\begin{subequations}
\begin{align}
(\epsilon+n \hbar \omega) a_l(m)=\hbar(-i k_y) b_l(m)-\frac{e v_F E_y}{\omega} b_{l-1}(m)\\
(\epsilon+n \hbar \omega) b_l(m)=\hbar(+i k_y) a_l(m)+\frac{e v_F E_y}{\omega} a_{l-1}(m)\\
(\epsilon+n \hbar \omega) c_l(m)=\hbar(-i k_y) d_l(m)-\frac{e v_F E_y}{\omega} d_{l-1}(m)\\
(\epsilon+n \hbar \omega) d_l(m)=\hbar(+i k_y) c_l(m)+\frac{e v_F E_y}{\omega} c_{l-1}(m)
\end{align}
\end{subequations}
For the lowest band in metallic acGNR, the energy of the carriers in the absence
of an applied electric field is $-\hbar v_F |k_y|$.
Following this procedure, we arrive at the following local current density terms defined in \cref{eq:jm}:
\begin{subequations}
\begin{align}
& j_{y}^{(1)}(m,\omega)=q v_F \left[i\left(a_1(m) b_0^{\dagger}(m) - a_0^{\dagger}(m) b_1(m) \right)+ i\left(c_1(m) d_0^{\dagger}(m) - c_0^{\dagger}(m) d_1(m) \right) \right]\\
& j_{ y}^{(3)}(m,\omega)=q v_F \left[i\left(a_2(m) b_1^{\dagger}(m) - a_1^{\dagger}(m) b_2(m) \right)+ i\left(c_2(m) d_1^{\dagger}(m) - c_1^{\dagger}(m) d_2(m) \right) \right]\\
& j_{ y}^{(3)}(m,3\omega)=q v_F \left[i\left(a_3(m) b_0^{\dagger}(m) -
a_0^{\dagger}(m) b_3(m) \right)+ i\left(c_3(m) d_0^{\dagger}(m) -
c_0^{\dagger}(m) d_3(m) \right) \right]\\
& j_{x}^{(1)}(m,\omega)=q v_F \left[\left(a_1(m) b_0^{\dagger}(m) + a_0^{\dagger}(m) b_1(m) \right)- \left(c_1(m) d_0^{\dagger}(m) + c_0^{\dagger}(m) d_1(m) \right) \right]\\
& j_{ x}^{(3)}(m,\omega)=q v_F \left[\left(a_2(m) b_1^{\dagger}(m) + a_1^{\dagger}(m) b_2(m) \right)- \left(c_2(m) d_1^{\dagger}(m) + c_1^{\dagger}(m) d_2(m) \right) \right]\\
& j_{ x}^{(3)}(m,3\omega)=q v_F \left[\left(a_3(m) b_0^{\dagger}(m) + a_0^{\dagger}(m) b_3(m) \right)- \left(c_3(m) d_0^{\dagger}(m) + c_0^{\dagger}(m) d_3(m) \right) \right]
\end{align}
\end{subequations}
We make the relaxation-free approximation, neglecting all scattering effects as
discussed above. We introduce an infinitesimal broadening factor \cite{wright09,ryu07,eo10,cheng1,*cheng1R,cheng2,cheng3,*cheng3R,mik16,scaling} $\Gamma$, by making the substitution $\omega=\omega+i \Gamma$ in the $\phi_{\mu}(m,l)$ spinor.
The \textit{i}th-order local nonlinear conductivity $\sigma_{\mu\nu}^{(i)}(m,\omega_0)$
is then obtained from \cref{eq:jm}
and summing over all states, with the Fermi energy $E_F$, $k_y= 2\pi m/L_y$ and
$\omega_y=v_F k_y$, we obtain the nonlinear conductance as:
\begin{equation}
\begin{aligned}\label{eq:gi}
g_{\mu \nu}^{(i)}(\omega_0) & =\lim_{\Gamma\to0} g_s g_v
\sum_{m=-\infty}^{\infty} \sigma_{\mu \nu}^{(i)}(m,\omega_0) N(\omega_y, E_F)\\
&=\lim_{\Gamma\to0} g_s g_v \frac{L_y}{2 \pi}\int_{-\infty}^{\infty} dk_y
\hspace{0.05cm} \sigma_{\mu \nu}^{(i)}(m,\omega_0) N(\omega_y, E_F)
\end{aligned}
\end{equation}
where the thermal factor in \cref{eq:gi} is:
\begin{equation}\label{eq:tfunc}
N(\omega_y, E_F) =n_F(-\hbar |\omega_y|, E_F) - n_F(\hbar |\omega_y|, E_F) =\frac{\sinh[\hbar |\omega_y|/(k_B T)]}{\cosh[E_F/(k_B T)]+\cosh[\hbar|\omega_y|/(k_B T)]}
\end{equation}

\section{Results and Discussion}\label{Results}
In what follows, we summarize the characteristics of the nonlinear conductance
for all combinations of applied field polarization and current direction.

\subsection{$E_x$}
If the applied electric field $\mathbf{E}$ is linearly polarized along the
transverse direction of the acGNR ($\hat {x}$ direction), for the metallic band where $k_{x,n}=0$, a net zero local current density for the $j_{x}(m,t)$ and $j_{y}(m,t)$ components is obtained. This result implies there
is neither linear nor third-order nonlinear current in metallic acGNR
when an electric field polarized transverse to the longitudinal direction of the
acGNR is applied.

\subsection{$E_y$}
For the case where the applied electric field $\mathbf{E}$ is linearly
polarized along the longitudinal
direction of the acGNR ($\hat {y}$ direction), for
metallic band where $k_{x,n}=0$, we arrive at the following expressions
for the isotropic nonlinear conductance:
\begin{equation}
\begin{aligned}\label{eq:gACy}
g_{yy}^{(1)}(\omega) & =g_0 \frac{g_s g_v v_F}{\omega L_x} \left[-
N(\frac{\omega}{2},E_F) \right] \\
g_{yy}^{(3)}(\omega) & =g_0 \frac{e^2 E_y^2 v_F^2}{\hbar ^2 \omega ^4} \frac{g_s
g_v v_F}{\omega L_x} \left[-2 N(\frac{\omega}{2},E_F) - N(\omega,E_F) \right]\\
g_{yy}^{(3)}(3\omega) & =g_0 \frac{e^2 E_y^2 v_F^2}{\hbar ^2 \omega ^4}
\frac{g_s g_v v_F}{\omega L_x} \left[\frac{1}{2} N(\frac{\omega}{2},E_F) -
N(\omega,E_F)+\frac{1}{2}N(\frac{3\omega}{2},E_F) \right]
\end{aligned}
\end{equation}
and the anisotropic nonlinear conductance:
\begin{equation}
\begin{aligned}\label{eq:gACx}
g_{yx}^{(1)}(\omega) & =g_0 \frac{g_s g_v v_F}{\omega L_x}
\left[N(\frac{\omega}{2},E_F) \right] \\
g_{yx}^{(3)}(\omega) & =g_0 \frac{e^2 E_y^2 v_F^2}{\hbar ^2 \omega ^4} \frac{g_s
g_v v_F}{\omega L_x} \left[ N(\omega,E_F) \right] \\
g_{yx}^{(3)}(3\omega) & =g_0 \frac{e^2 E_y^2 v_F^2}{\hbar ^2 \omega ^4}
\frac{g_s g_v v_F}{\omega L_x} \left[-\frac{1}{2} N(\frac{\omega}{2},E_F) +
N(\omega,E_F)-\frac{1}{2}N(\frac{3\omega}{2},E_F) \right]
\end{aligned}
\end{equation}
with the $N(\omega)$ defined in \cref{eq:tfunc},
and the quantum conductance $g_0=\frac{e^2}{4\hbar}$. Due to the inversion
symmetry inherent in acGNR, the \textit{2}nd-order current makes no contribution to the total current.\par

In the discussion below, we compare our results for the nonlinear conductance of
metallic acGNR with those reported by Wright, \textit{et.al.}\cite{wright09} and
Ang \textit{et.al.} \cite{ang2015nonlinear} for intrinsic 2D SLG.
In Eq. 70 of Ang \textit{et.al.} \cite{ang2015nonlinear}, they write the
expression for the third-order Kerr conductance as:
\begin{equation}
g^{(3)}(\omega)_{2D}=-g_0 \frac{e^2 E_0^2 v_F^2}{\hbar ^2 \omega ^4} \left[2
\tanh\left(\frac{\hbar \omega}{2 k_B T}\right)\right]\label{2Dg3w1v1}
\end{equation}
We believe this expression omits an additional required term due to the
resonance at $\epsilon = \hbar \omega/2$. The correct expression for the
third-order Kerr conductance is:
\begin{equation}\label{2Dg3w1v2}
g^{(3)}(\omega)_{2D}=-g_0 \frac{e^2 E_0^2 v_F^2}{\hbar ^2 \omega ^4} \left[
\frac{5}{4}N\left(\frac{\omega}{2},E_F \right)+2N(\omega, E_F) \right]
\end{equation}
Notice that for intrinsic 2D SLG ($E_F=0$),
$N(\omega,0)=\tanh\left[\hbar|\omega|/(2k_B T)\right]$, and we recover the thermal factor used in Refs. \cite{wright09,ang2015nonlinear}.
The missing $ \frac{5}{4}N(\frac{\omega}{2})$ term  in Eq.
\ref{2Dg3w1v1} is the missing contribution for $|\epsilon|=\hbar
\omega/2$. As both $\epsilon=\pm \hbar \omega/2$ and $\epsilon=\pm \hbar
\omega$ contribute to the generation of the third-order Kerr current
\cite{cheng1,*cheng1R,cheng3,*cheng3R,mik16}, we believe
that Eq. \ref{2Dg3w1v2} is correct. At $T = 0 \, \mathrm{K}$, the real part of
the Kerr conductance has two threshold frequencies, $\omega=\pm 2 E_F/\hbar$ and
$\omega=\pm E_F/\hbar$, corresponding to the contribution for states
with energies
$\epsilon=\pm \hbar\omega/2$ and $\pm \epsilon=\hbar\omega$, or the resonant transitions
for which the Fermi level gap $2|E_F/\hbar|$ matches the one photon and two photon
frequencies respectively \cite{cheng1,*cheng1R,cheng3,*cheng1R}. We note that
the zero temperature result of Refs.
\cite{mik14,*mik14R,mik16,cheng1,*cheng1R,cheng2,cheng3,*cheng3R,scaling}
contain the same threshold frequencies.
As a result, the $N$-photon coupling
approach we have adopted \cite{wright09,ang2015nonlinear} here and the quantum
theories of the third-order nonlinear response
\cite{mik14,*mik14R,mik16,cheng1,*cheng1R,cheng2,cheng3,*cheng3R,scaling} show
qualitative agreement. The position of the peaks shown in the
plots of Refs\cite{mik16,cheng1,cheng2,cheng3} in the absence of broadening
are at the threshold frequencies with respect to $E_F/\hbar$ derived from
\cref{2Dg3w1v2} at $T = \SI{0}{K}$. Therefore, we compute $g^{(3)}(\omega)$ for 2D SLG using \cref{2Dg3w1v2} in what follows.

In Eq. (71) of Ang \textit{et.al.} \cite{ang2015nonlinear}, they write the
expression for the third-order third-harmonic conductance as:
\begin{equation}
g^{(3)}(3\omega)_{2D}=g_0 \frac{e^2 E_0^2 v_F^2}{\hbar ^2 \omega ^4} \left[
\frac{13}{48}\tanh(\frac{\hbar \omega}{4 k_B
T})-\frac{2}{3}\tanh(\frac{\hbar \omega}{2 k_B T})+
\frac{45}{48}\tanh(\frac{3\hbar \omega}{4 k_B T})\right]\label{2Dg3w3v1}
\end{equation}
Our analysis of the problem gives the same set of coefficients as 
\cref{2Dg3w3v1}, to wit:
\begin{equation}\label{2Dg3w3v2}
g^{(3)}(3\omega)_{2D}=g_0 \frac{e^2 E_0^2 v_F^2}{\hbar ^2 \omega ^4} \left[
\frac{13}{48}N(\frac{\omega}{2}, E_F)-\frac{2}{3}N(\omega, E_F)+
\frac{45}{48}N(\frac{3\omega}{2},E_F)\right]
\end{equation}
For intrinsic 2D SLG ($E_F=0$), $N(\omega,0)=\tanh\left[\hbar|\omega|/(2k_B
T)\right]$, and therefore \cref{2Dg3w3v2} reduces to \cref{2Dg3w3v1} used in Refs. \cite{wright09,ang2015nonlinear}.
As a result, we compute $g^{(3)}(3 \omega)$ for intrinsic 2D SLG using 
\cref{2Dg3w3v2} in what follows. The three threshold frequencies in \cref{2Dg3w3v2}
are the same as those obtained by Morimoto \textit{et.al.} \cite{scaling}. At $T
= \SI{0}{K}$, the resonant frequencies are $\omega=\pm 2E_F/\hbar$,
$\omega=\pm E_F/\hbar$ and $\omega=\pm 2E_F/3\hbar$, corresponding to the contribution
for states at $\epsilon=\pm \hbar\omega/2$, $\epsilon=\pm \hbar\omega$ and
$\epsilon=\pm 3\hbar\omega/2$, or the resonant transitions for which the Fermi
level gap $2|E_F/\hbar|$ matches the frequencies of the one photon, two photon,
and three photon transitions respectively
\cite{cheng1,*cheng1R,cheng3,*cheng3R}. Interestingly, the coefficients for
$\omega/2$, $\omega$ and $3\omega/2$ for the third-harmonic expression in Refs.
\cite{cheng1,*cheng1R,cheng2,cheng3,*cheng3R}, are $17/48$, $-4/3$ and $45/48$
respectively. As Mikhailov pointed out, different theories of the \si{THz}
nonlinear response in 2D SLG may show somewhat contradictory \cite{mik16}
results, the difference between these coefficients being due to the extreme
complexity of the problem.  However, we point out that \cref{2Dg3w3v2} shows
that the main contribution for third-harmonic conductance is from the
$3\omega/2$ resonance. This observation is confirmed by the results from three
independent models: Wright \textit{et.al.} \cite{wright09}, Mikhailov \cite{mik16} and Cheng \textit{et.al.} \cite{cheng1,*cheng1R,cheng2,cheng3,*cheng3R}. 

A thorough analysis of our objection to the Wright
\textit{et.al.}\cite{wright09}
and Ang \textit{et.al.}\cite{ang2015nonlinear} calculation of the Kerr
conductance for intrinsic 2D SLG is provided in the Appendix below.\par

The total third-order nonlinear current for metallic acGNR can be
expressed as:
\begin{equation}\label{eq:g3yy}
J_{\nu}^{(3)}(t) =g_{y \nu}^{(3)}(\omega) E_y e^{-i \omega t}+g_{y
\nu}^{(3)}(3\omega) E_y e^{-i 3 \omega t} + c.c. 
\end{equation}
This result shows that for metallic acGNR, the third-order
nonlinear current is a superposition of two frequency terms:  (i)
$g_{y\nu}^{(3)}(\omega)$, the third-order Kerr conductance, which has a single
frequency electron current density term
corresponding to the absorption of two photons and the simultaneous emission of
one photon; and (ii), $g_{y\nu}^{(3)}(3 \omega)$, the third-order third-harmonic conductance
term corresponding to the simultaneous absorption of three
photons. The complex conjugate parts in Eq. \ref{eq:g3yy} are for
the emission process.

In this paper we consider the case where the length of the ribbon $L_y
\rightarrow \infty$,
and as a result, we have a
quasi continuum of states for the linear bands near the Dirac points in
metallic acGNR. To simplify the discussion, we present results for acGNR20,
the armchair graphene nanoribbon $N=20$ atoms wide.

\begin{figure}[H]
\centering
\subfloat{\label{fig:2a}\includegraphics[width=0.5\linewidth]{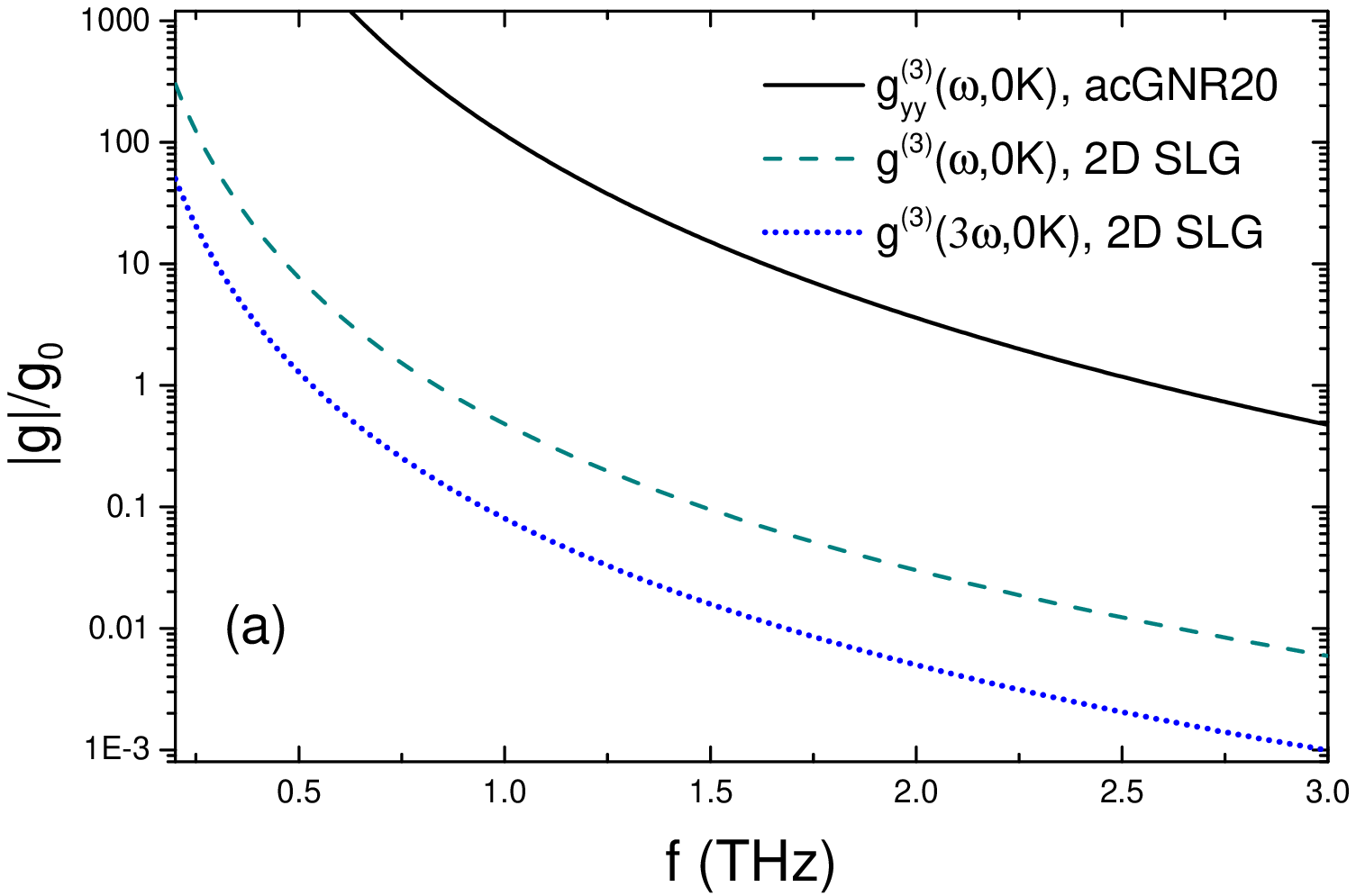}}
\hfil
\subfloat{\label{fig:2b}\includegraphics[width=0.5\linewidth]{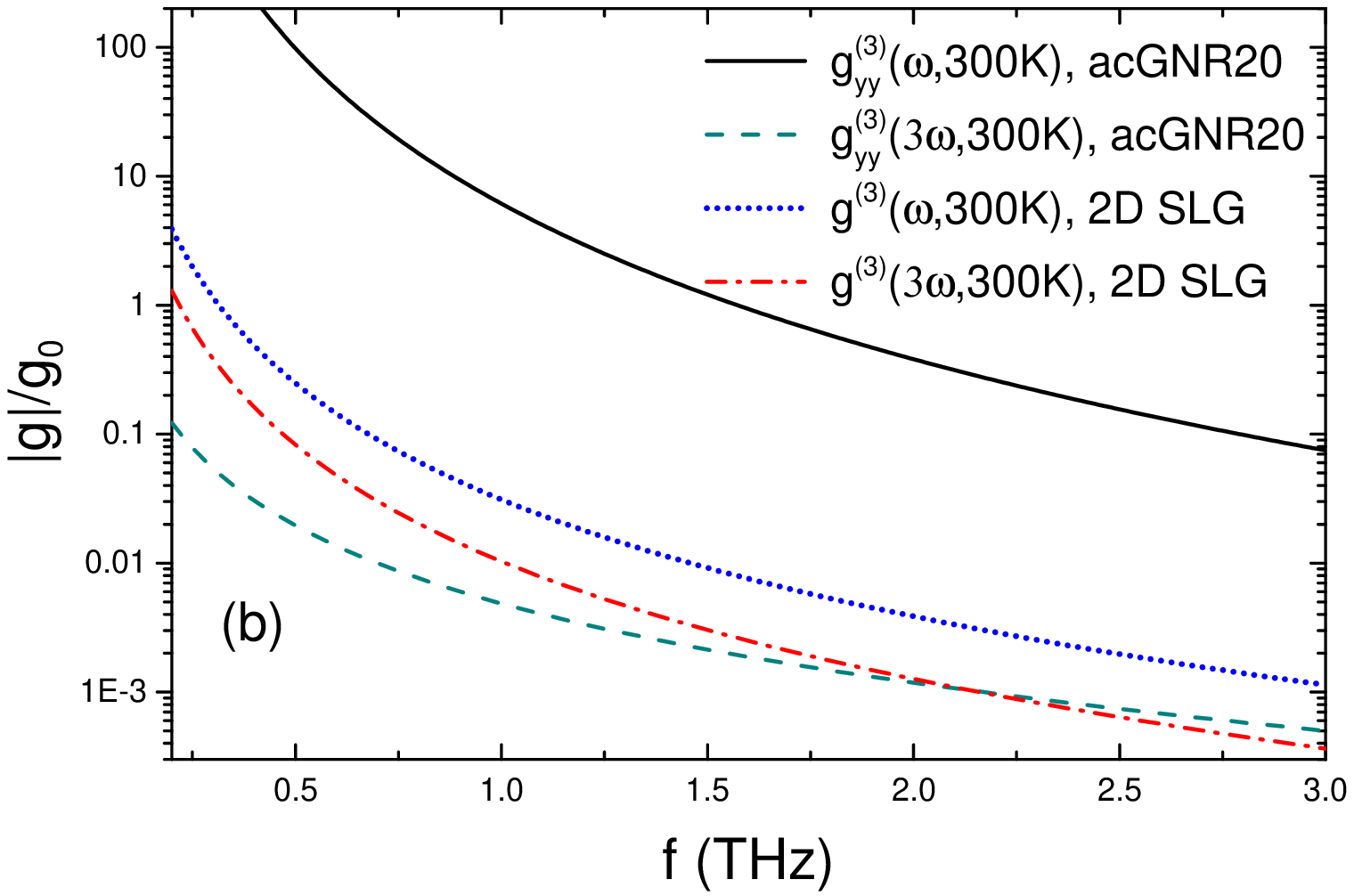}}
\hfil
\subfloat{\label{fig:2c}\includegraphics[width=0.5\linewidth]{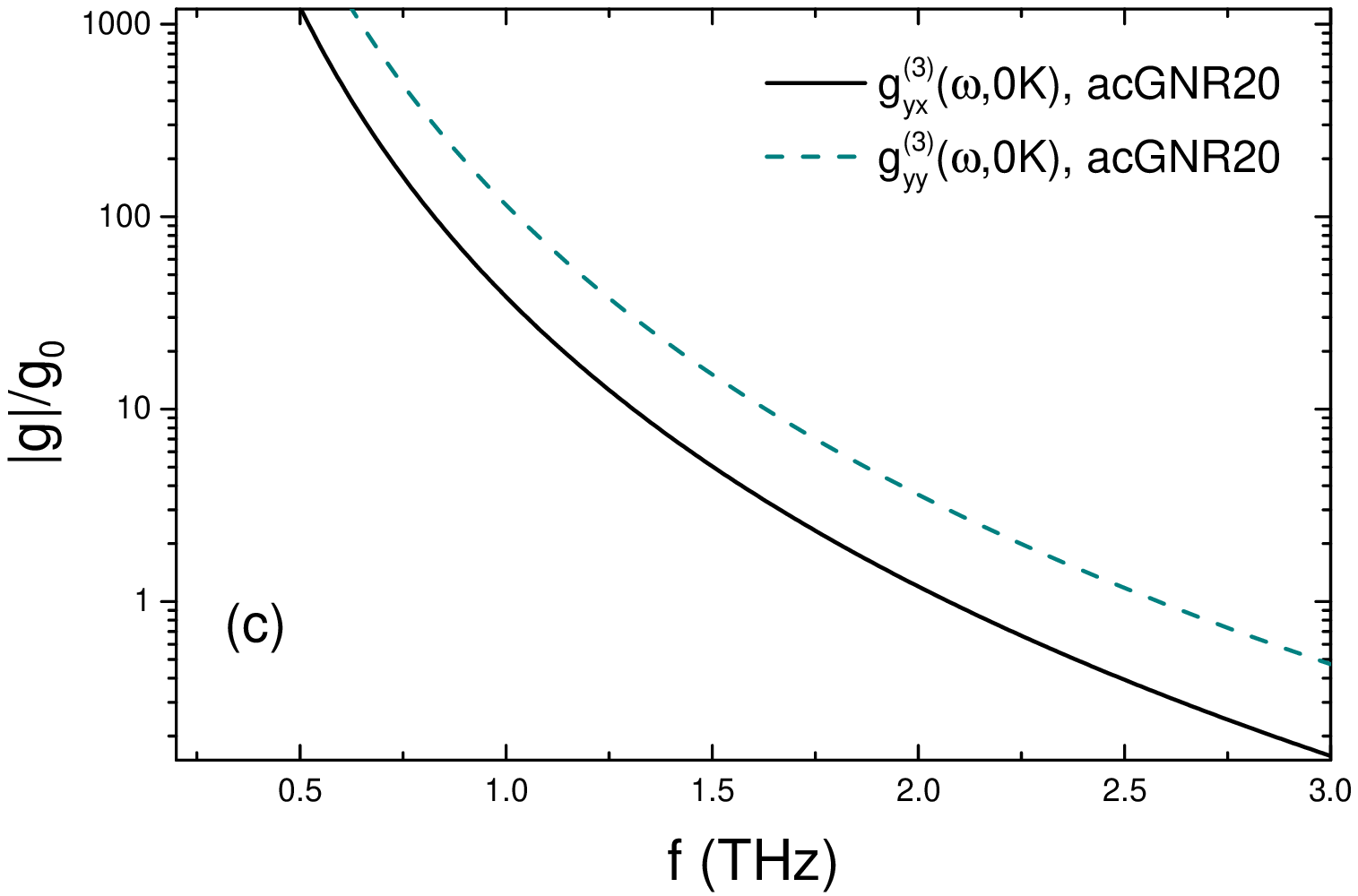}}
\hfil
\subfloat{\label{fig:2d}\includegraphics[width=0.5\linewidth]{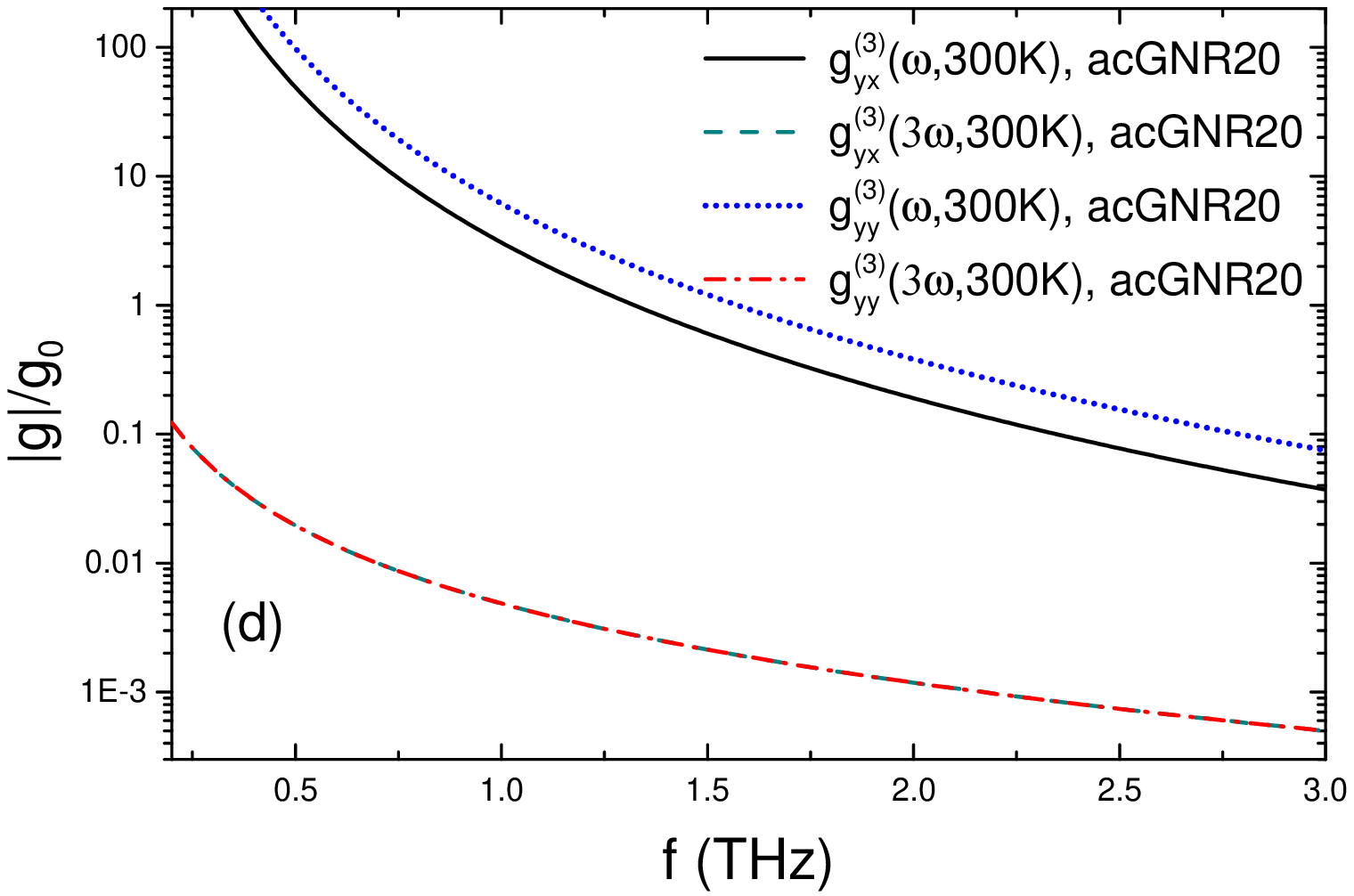}}
\caption{(color online) Comparison of the Kerr and third-harmonic nonlinear conductances
for intrinsic acGNR20 with 2D SLG at (a) $T = 0
\, \mathrm{K}$ and (b) $T = 300 \, \mathrm{K}$; and comparison of
isotropic and anisotropic conductances for acGNR20 at (c) $T = 0 \, \mathrm{K}$ and
(d) $T = 300 \, \mathrm{K}$. The field strength used in all calculations is
$E_y = 10 \, \mathrm{kV/m}$ and the excitation frequency $f = \omega/2 \pi$.}
\label{fig:2}
\end{figure}

\begin{figure}[H]
\centering
\subfloat{\label{fig:3a}\includegraphics[width=0.5\linewidth]{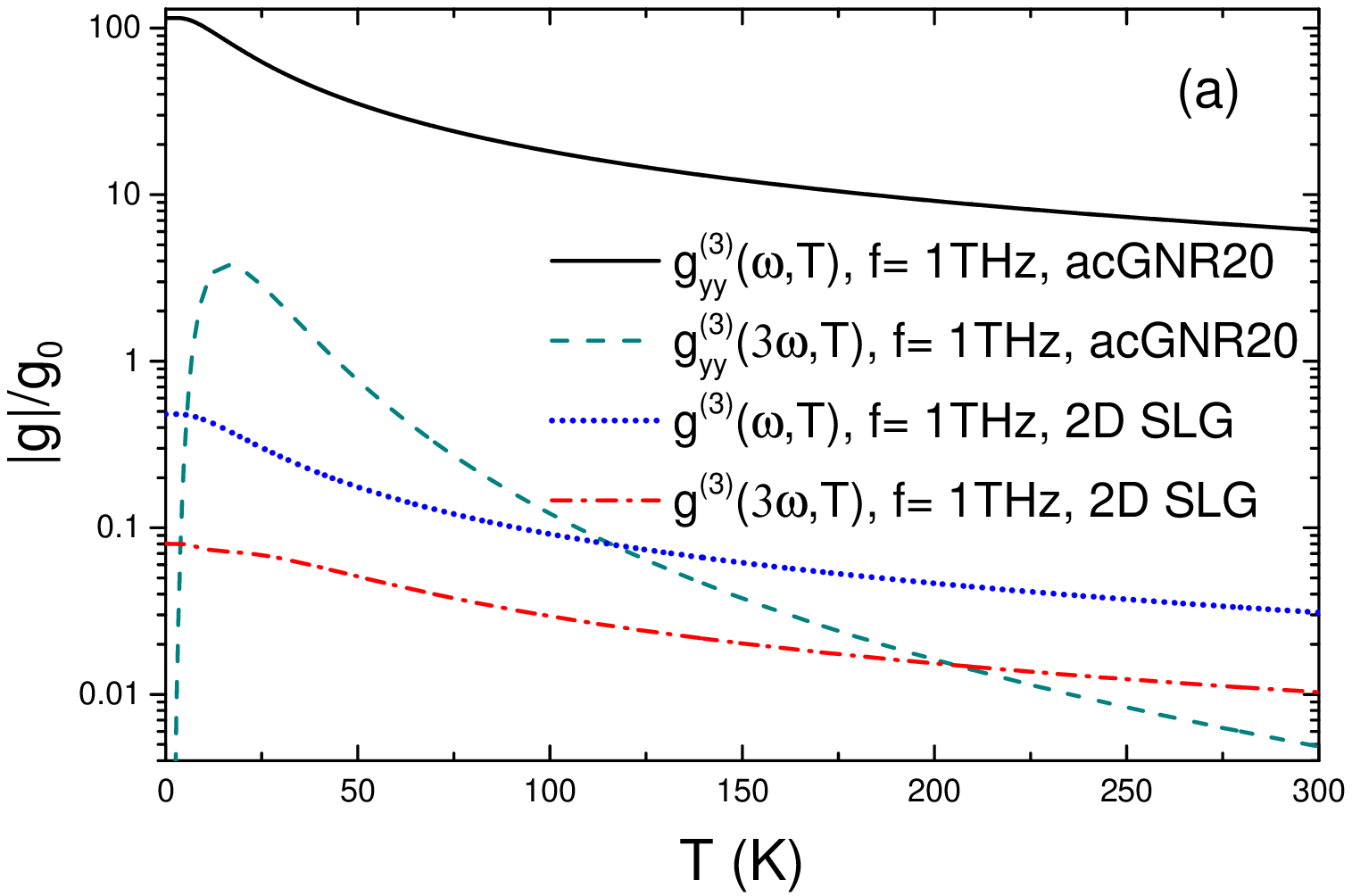}}
\hfil
\subfloat{\label{fig:3b}\includegraphics[width=0.5\linewidth]{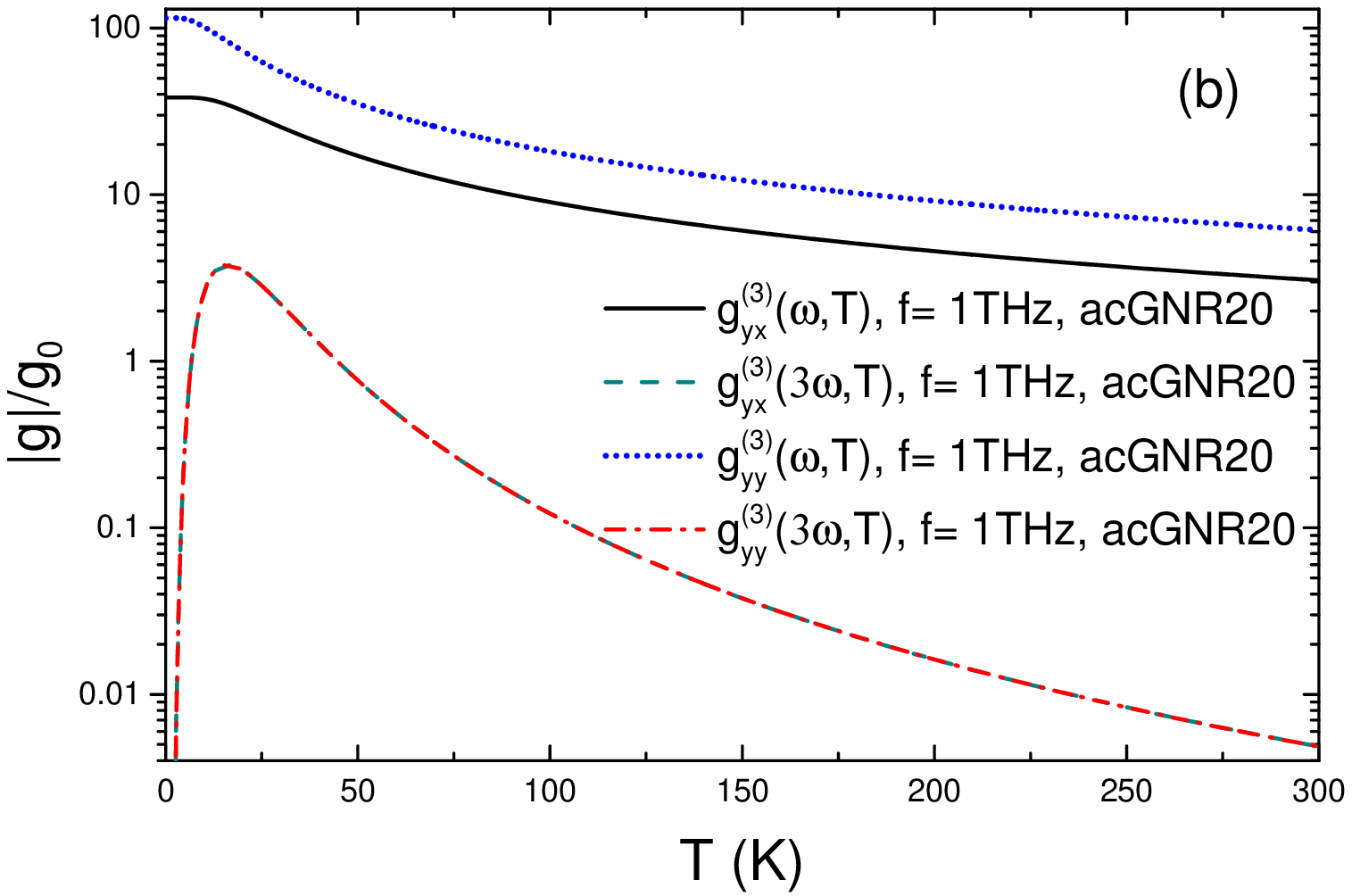}}
\hfil
\subfloat{\label{fig:3c}\includegraphics[width=0.5\linewidth]{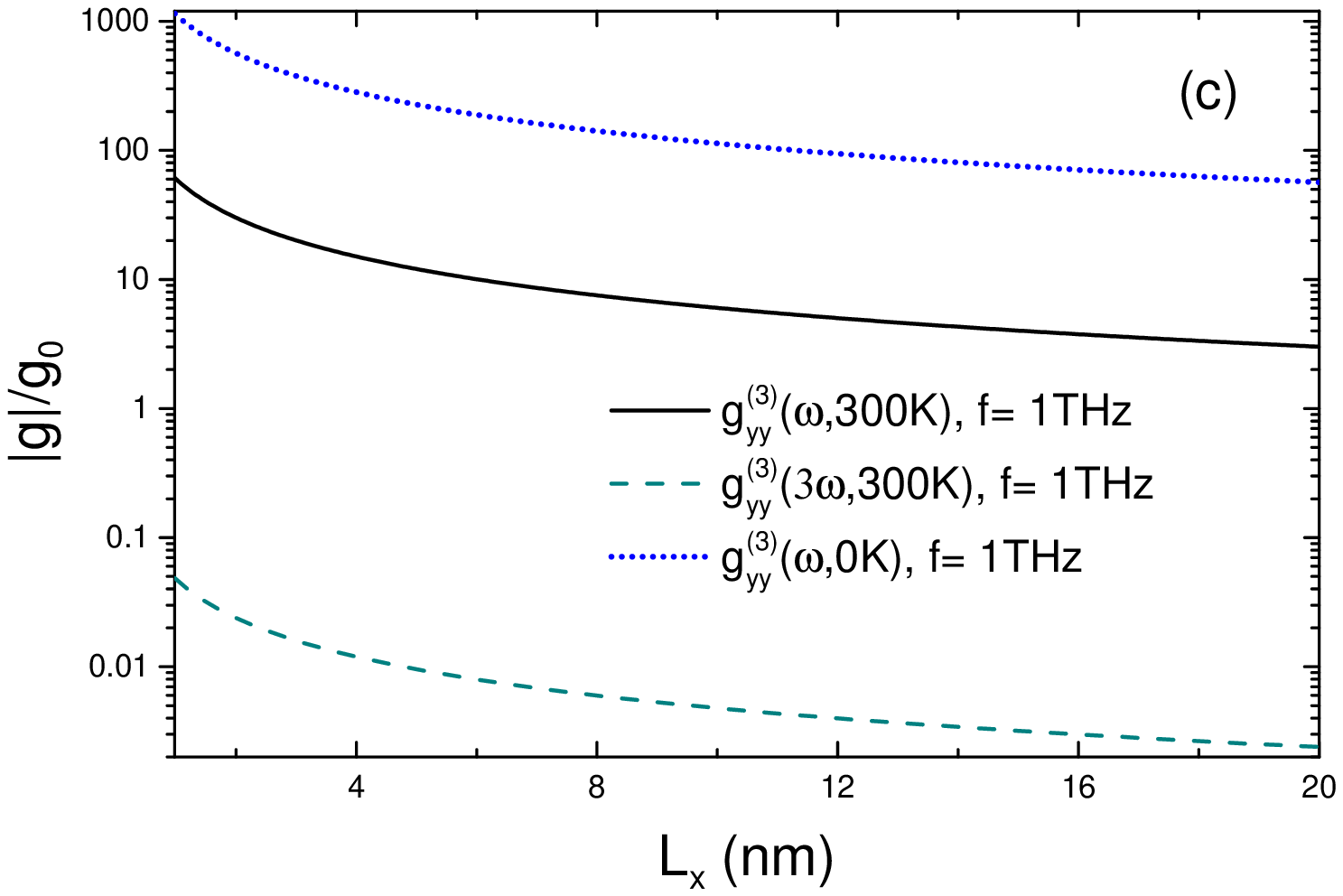}}
\hfil
\subfloat{\label{fig:3d}\includegraphics[width=0.5\linewidth]{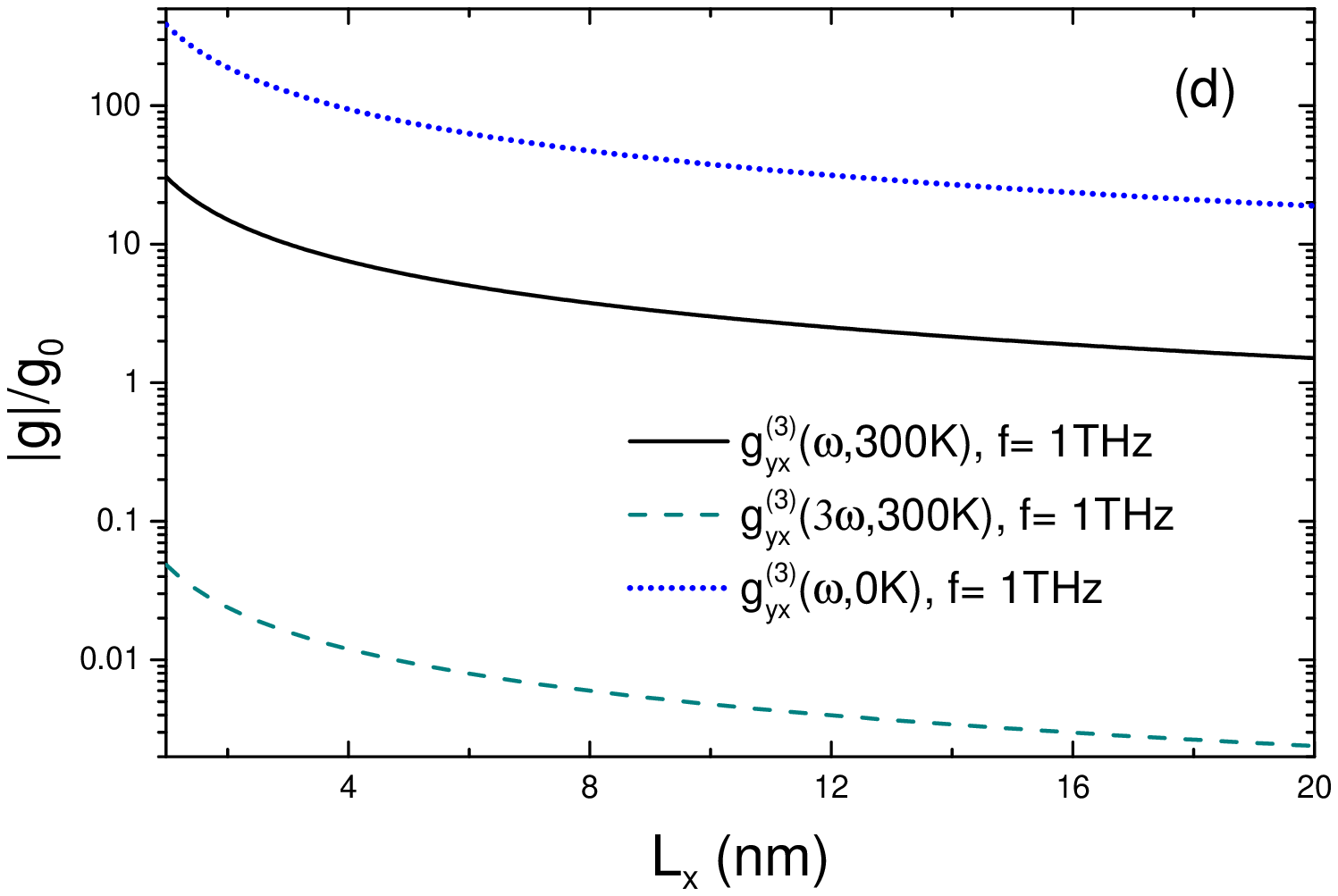}}
\caption{(color online) Comparison of the temperature dependence of the Kerr and
third-harmonic nonlinear conductances
for (a) intrinsic acGNR20 with that of 2D SLG;  
(b) isotropic and anisotropic nonlinear conductances
for intrinsic acGNR20; comparison of the nanoribbon width dependence of (c)
the Kerr and third-harmonic isotropic nonlinear conductances; and (d)
the Kerr and third-harmonic anisotropic nonlinear conductance. The excitation
frequency used in all calculations is $f = \omega/2 \pi = 1 \, \mathrm{THz}$.}
\label{fig:3}
\end{figure}

\begin{figure}[H]
\centering
\subfloat{\label{fig:4a}\includegraphics[width=0.5\linewidth]{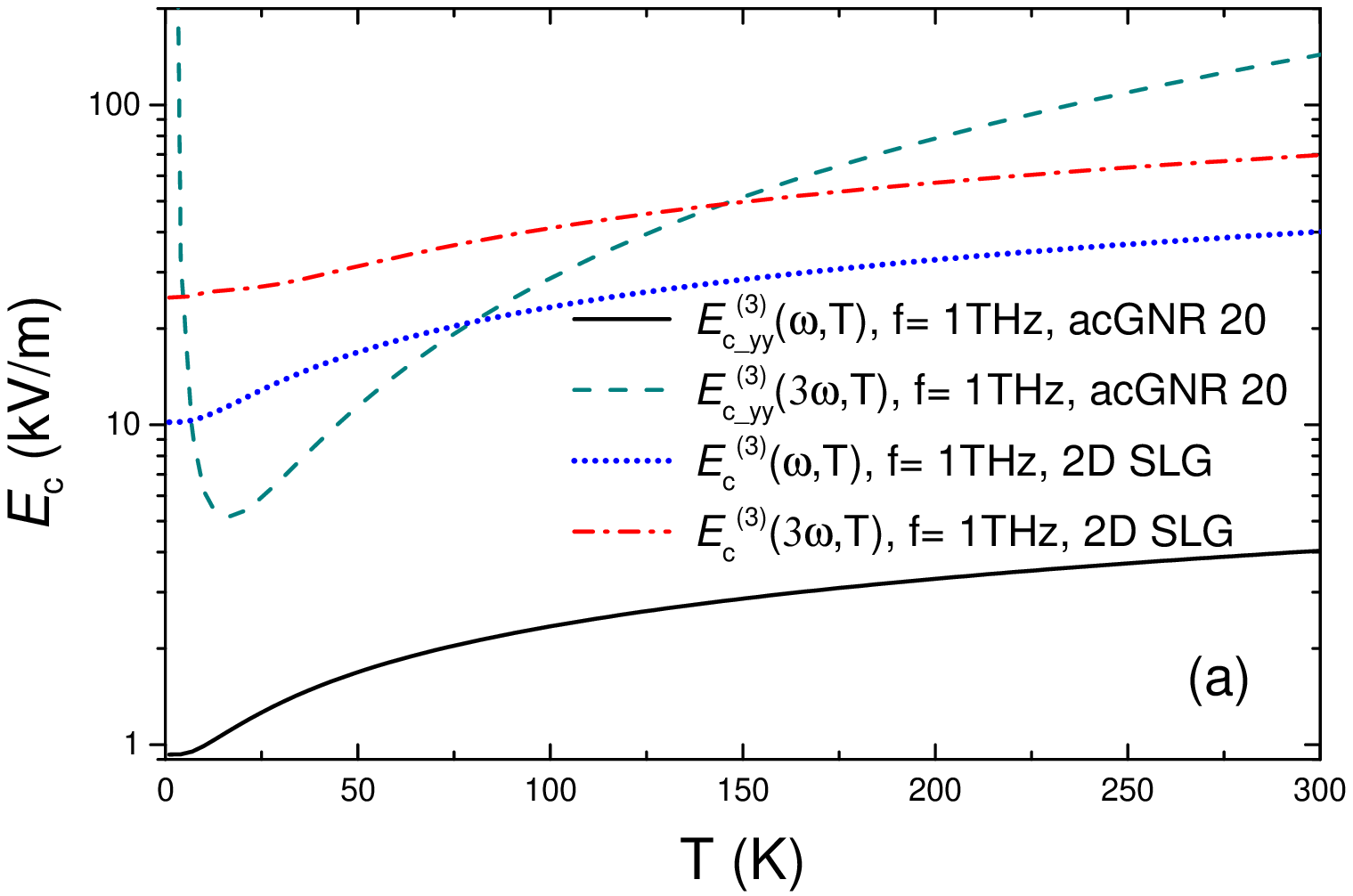}}
\vfil
\subfloat{\label{fig:4b}\includegraphics[width=0.5\linewidth]{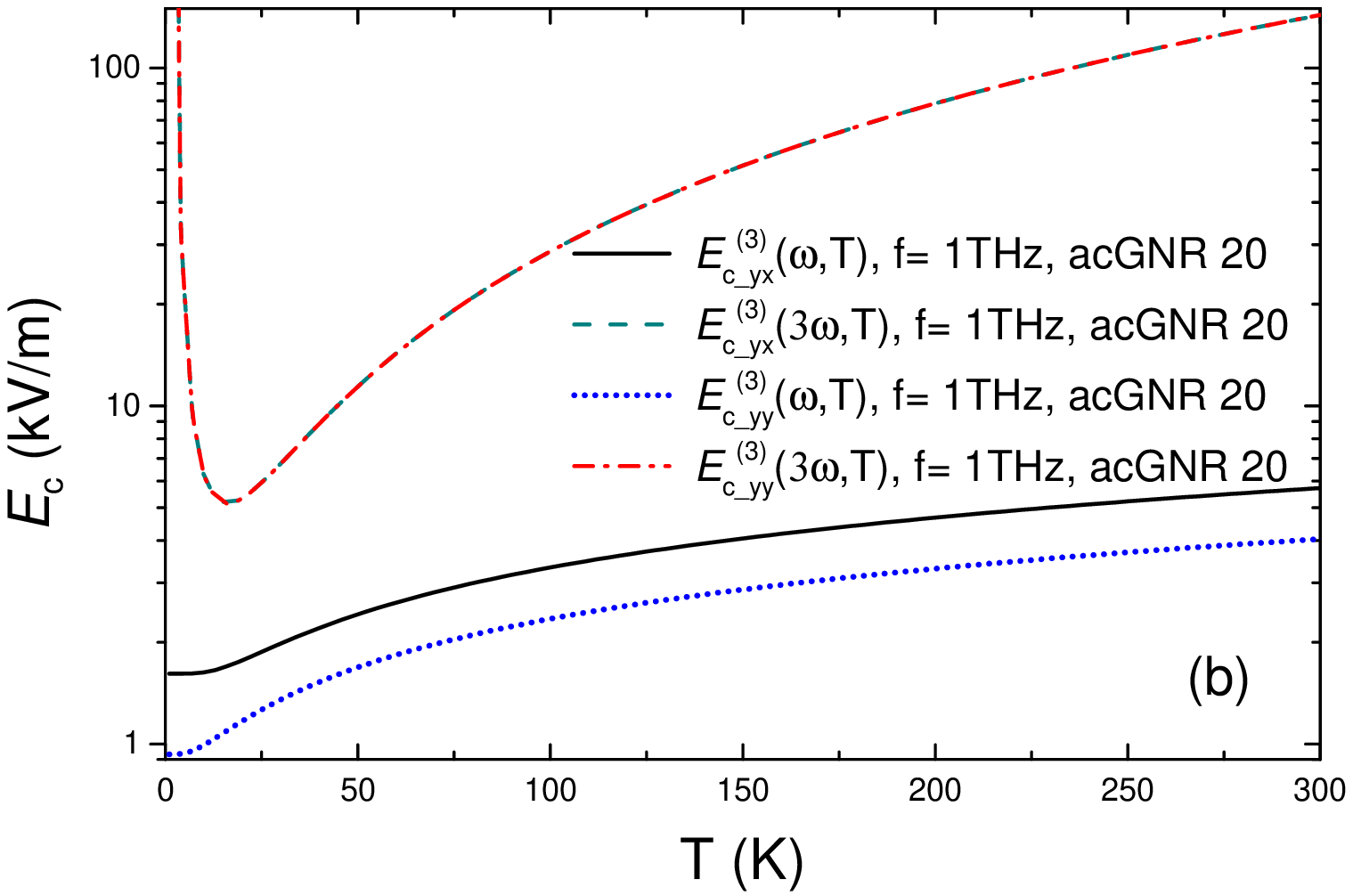}}
\caption{(color online) Comparison of the temperature dependence of the critical fields
for (a) the isotropic Kerr and third-harmonic processes for intrinsic acGNR20
with those of 2D SLG; and (b) the isotropic and anisotropic Kerr and
third-harmonic nonlinear processes for intrinsic acGNR20. The excitation
frequency used in all calculations is $f = \omega/2 \pi = 1 \, \mathrm{THz}$.}
\label{fig:4}
\end{figure}

\begin{figure}[H]
\centering
\subfloat{\label{fig:5a}\includegraphics[width=0.5\linewidth]{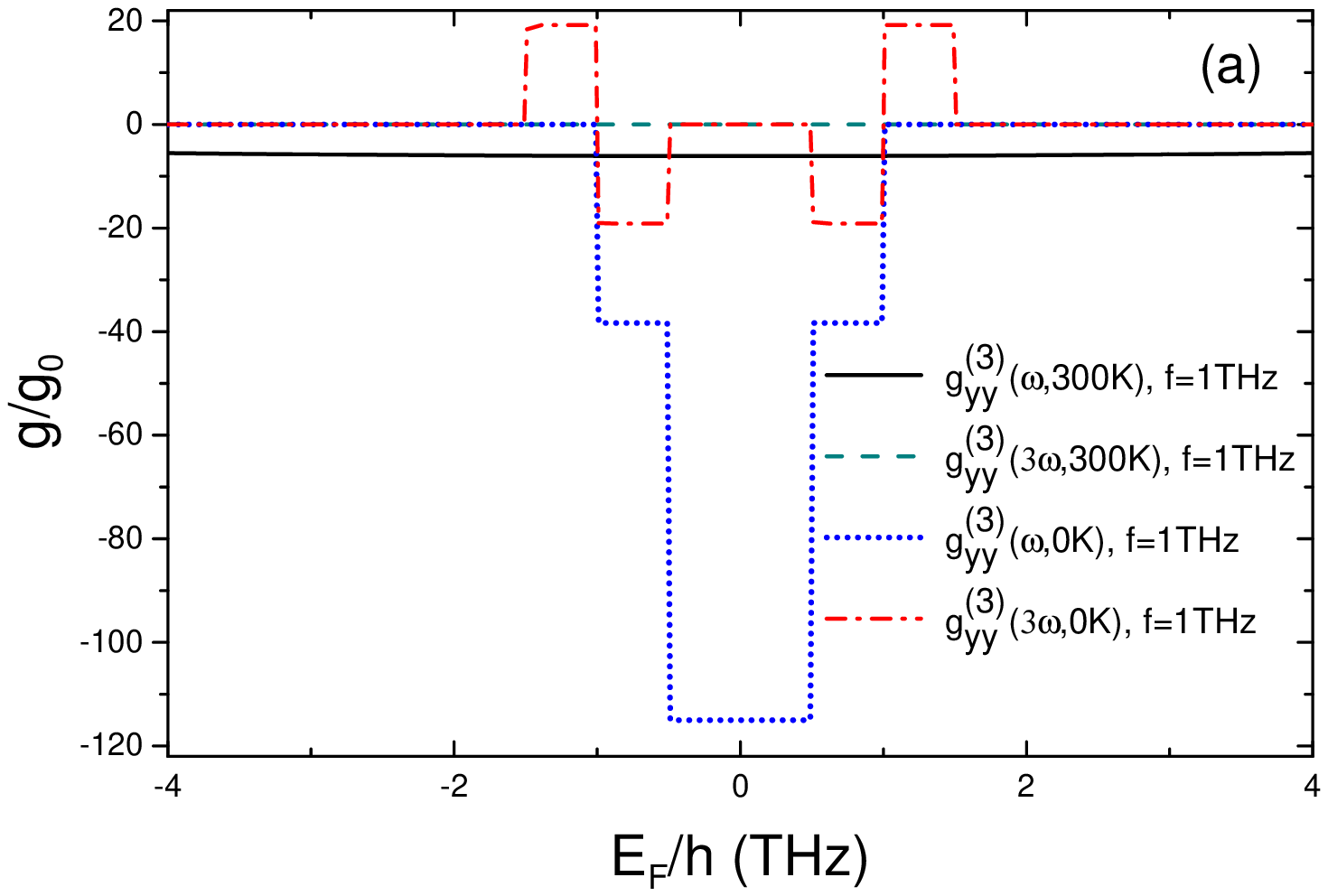}}
\vfil
\subfloat{\label{fig:5b}\includegraphics[width=0.5\linewidth]{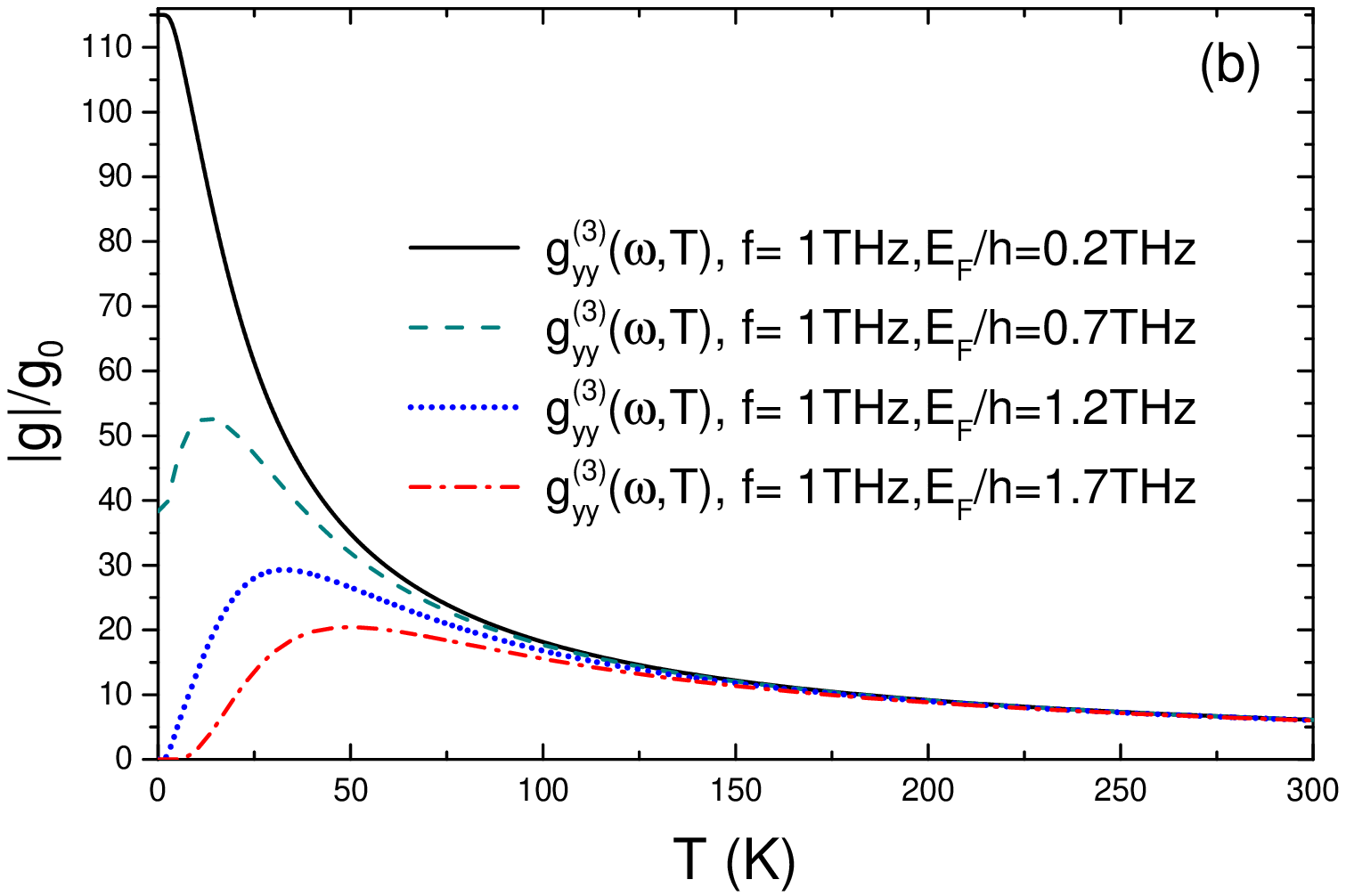}}
\vfil
\subfloat{\label{fig:5c}\includegraphics[width=0.5\linewidth]{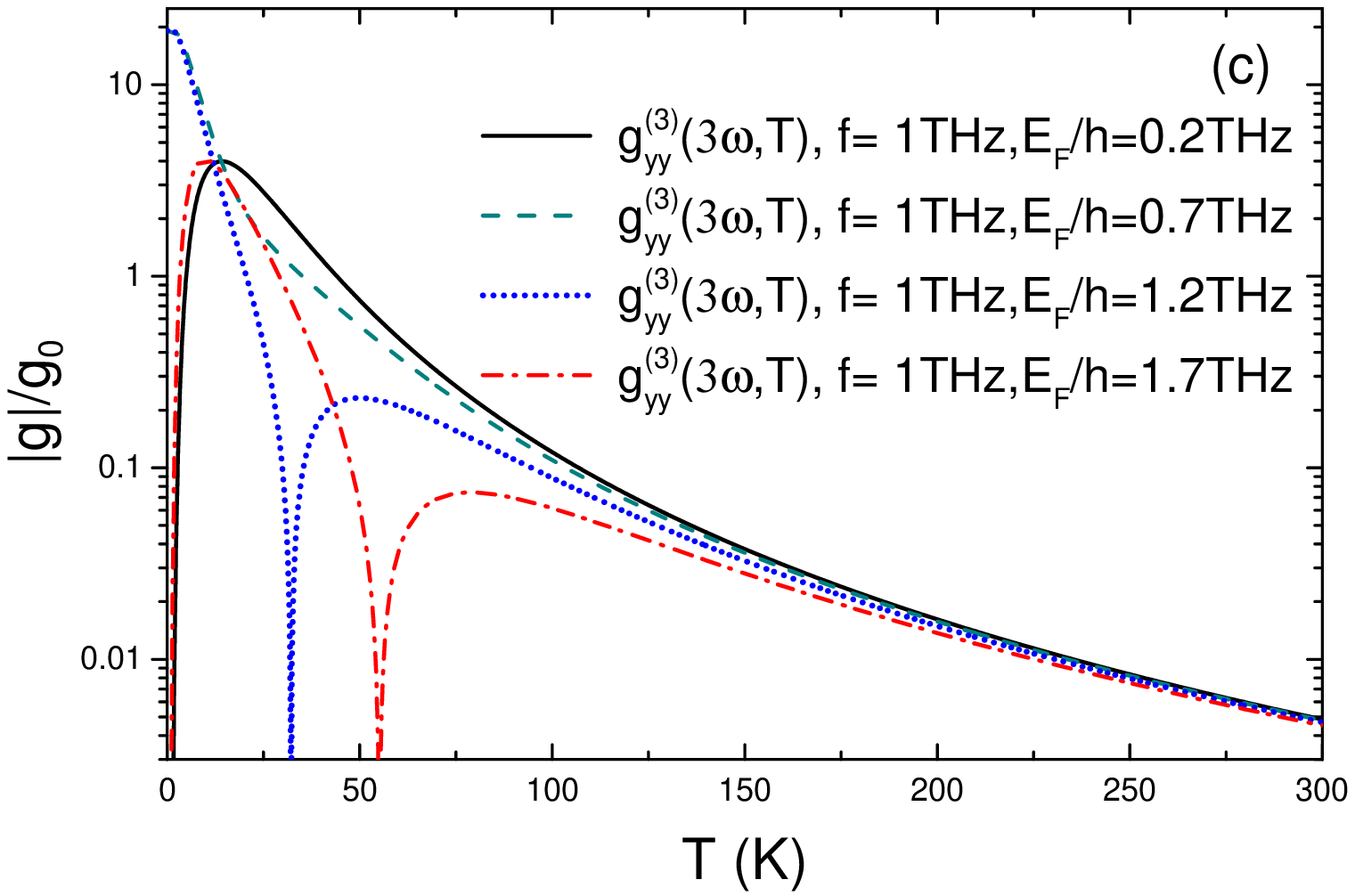}}
\caption{(color online) (a) The $E_F$ dependence of the isotropic Kerr and third-order
nonlinear conductances of acGNR20 at $T = 0 \, \mathrm{K}$ and $T = 300 \, \mathrm{K}$;
(b) the temperature dependence of the isotropic Kerr nonlinear conductance of
acGNR20 for
various Fermi levels; and (c) the temperature dependence of the isotropic
third-harmonic nonlinear conductances of acGNR20 for various Fermi levels. The excitation
frequency used in all calculations is $f = \omega/2 \pi = 1 \, \mathrm{THz}$.}
\label{fig:5}
\end{figure}

\begin{figure}[H]
\centering
\subfloat{\label{fig:6a}\includegraphics[width=0.5\linewidth]{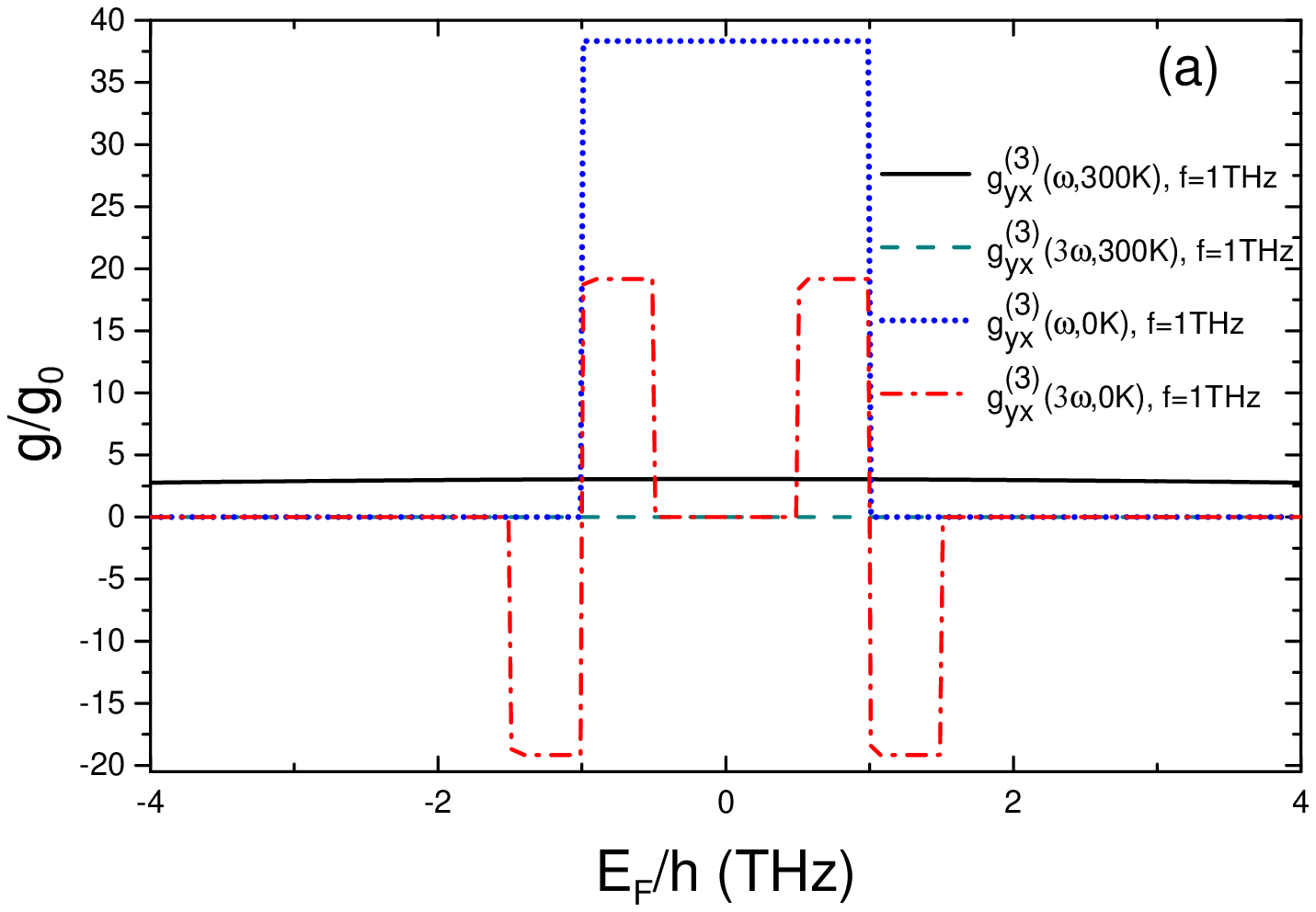}}
\vfil
\subfloat{\label{fig:6b}\includegraphics[width=0.5\linewidth]{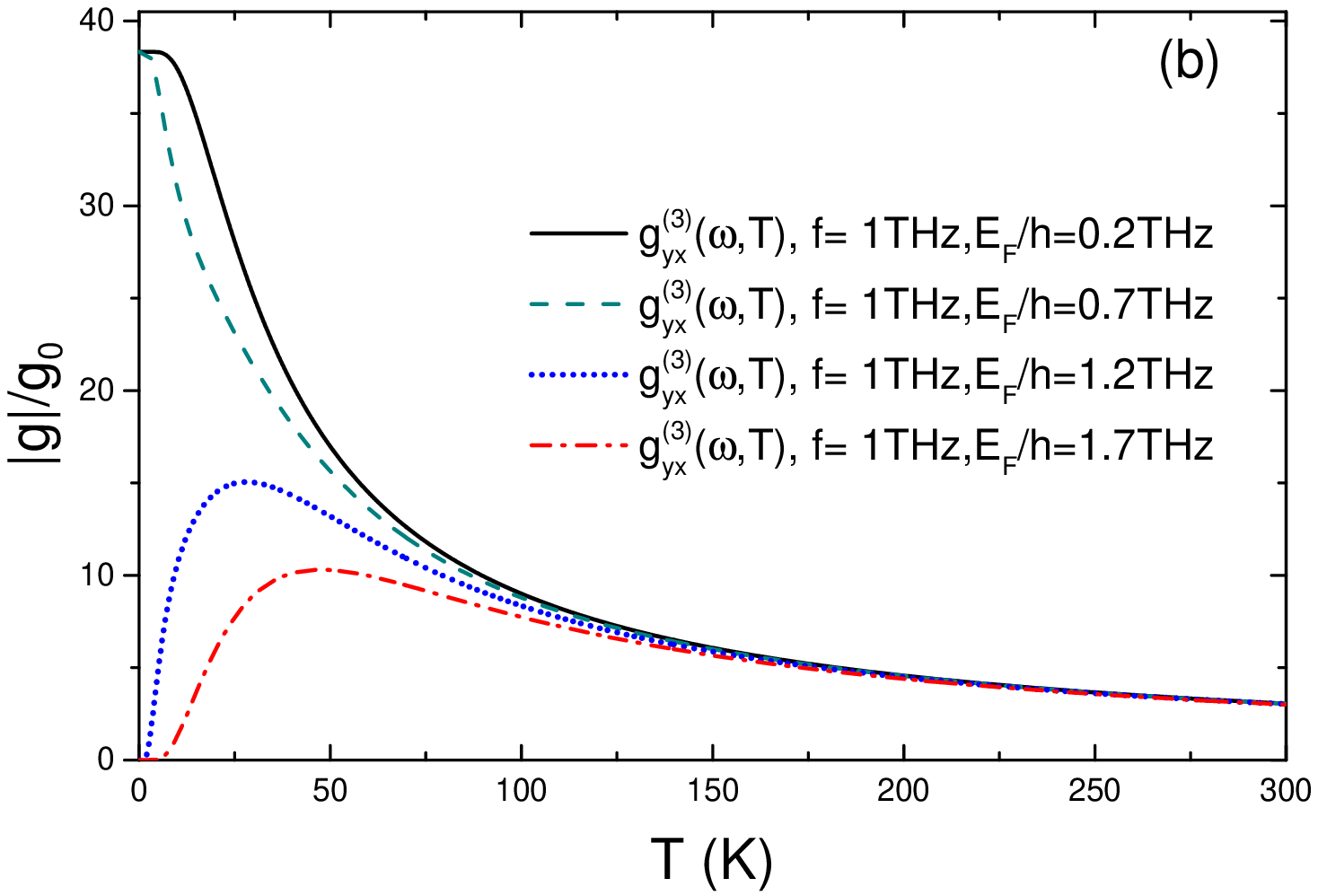}}
\vfil
\subfloat{\label{fig:6c}\includegraphics[width=0.5\linewidth]{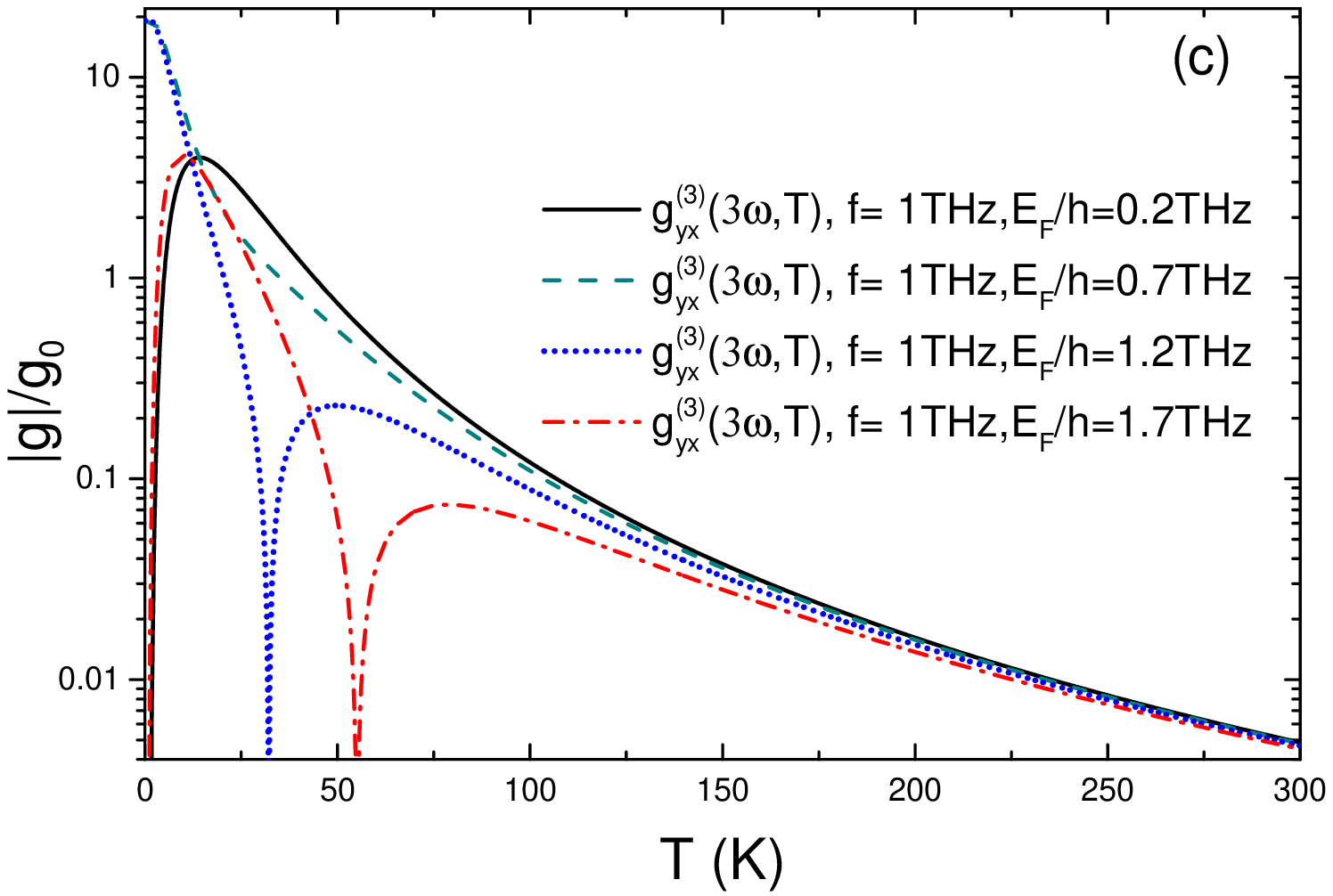}}
\caption{(color online) (a) The $E_F$ dependence of the anisotropic Kerr and third-order
nonlinear conductances of acGNR20 at $T = 0 \, \mathrm{K}$ and $T = 300 \, \mathrm{K}$;
(b) the temperature dependence of the anisotropic Kerr nonlinear conductances of
acGNR20 for various Fermi levels; and (c) the temperature dependence of the anisotropic
third-harmonic nonlinear conductances of acGNR20 for various Fermi levels. The excitation
frequency used in all calculations is $f = \omega/2 \pi = 1 \, \mathrm{THz}$.}
\label{fig:6}
\end{figure}

\begin{figure}[H]
\centering
\subfloat{\label{fig:7a}\includegraphics[width=0.5\linewidth]{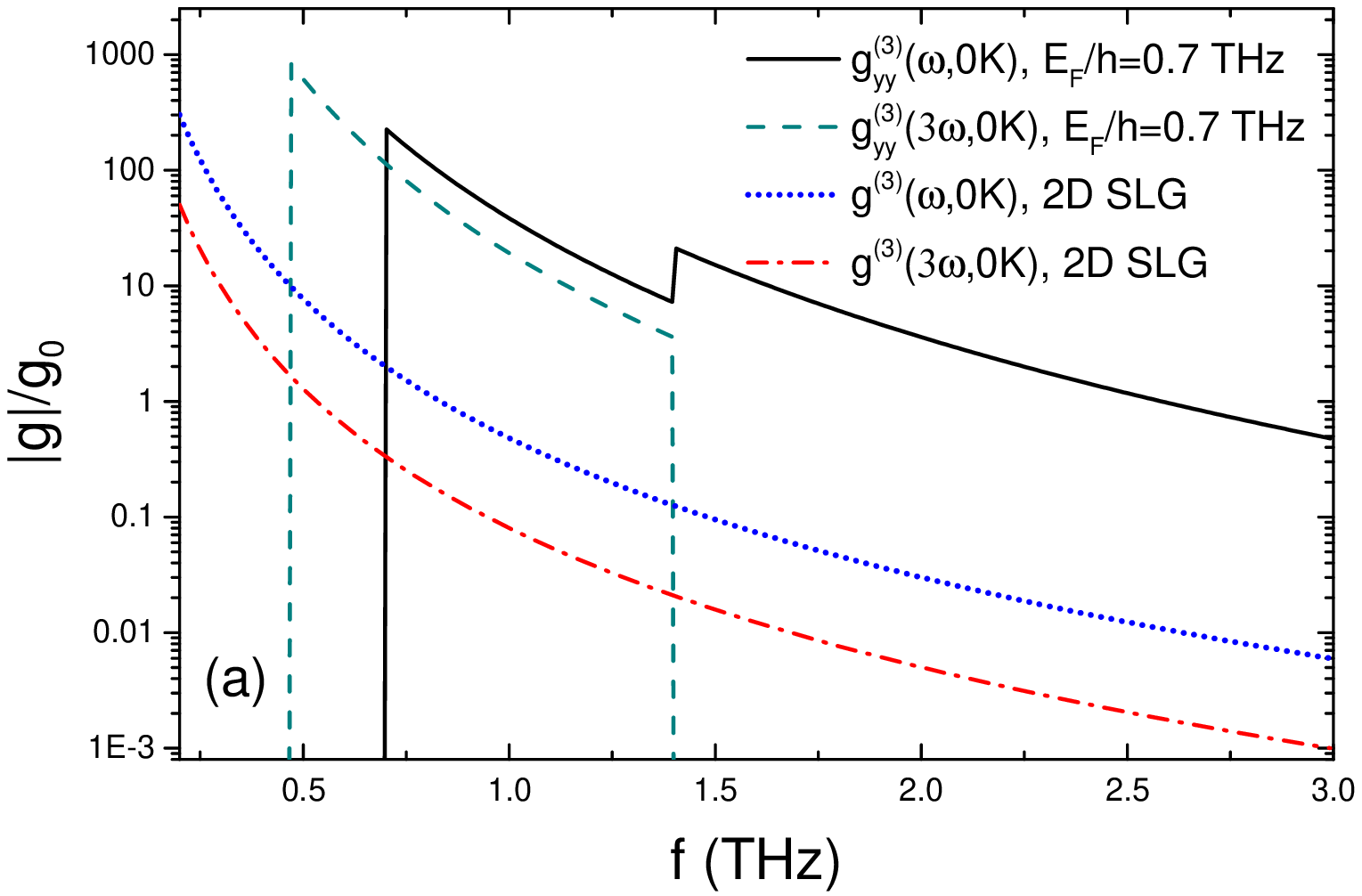}}
\hfil
\subfloat{\label{fig:7b}\includegraphics[width=0.5\linewidth]{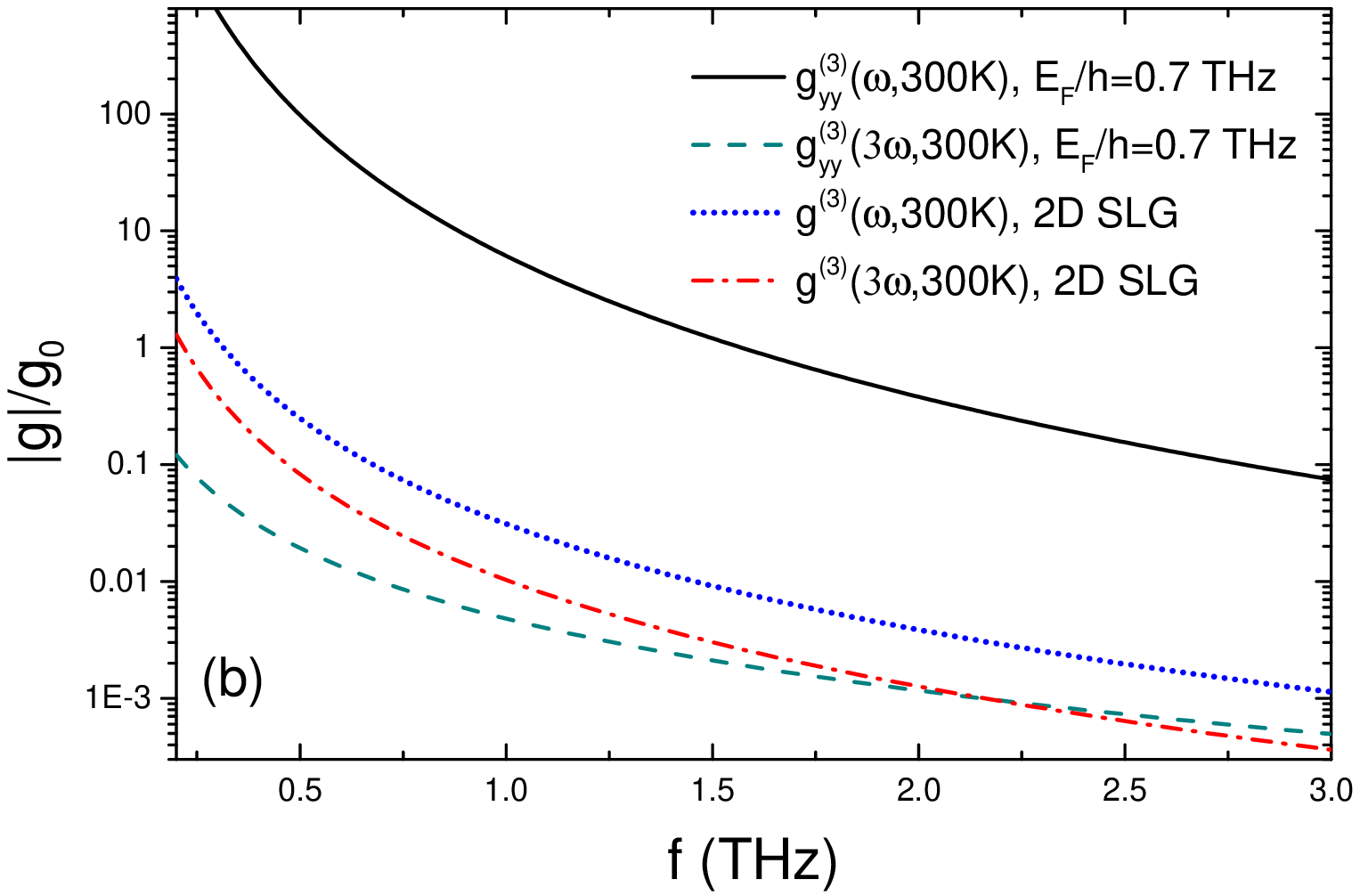}}
\hfil
\subfloat{\label{fig:7c}\includegraphics[width=0.5\linewidth]{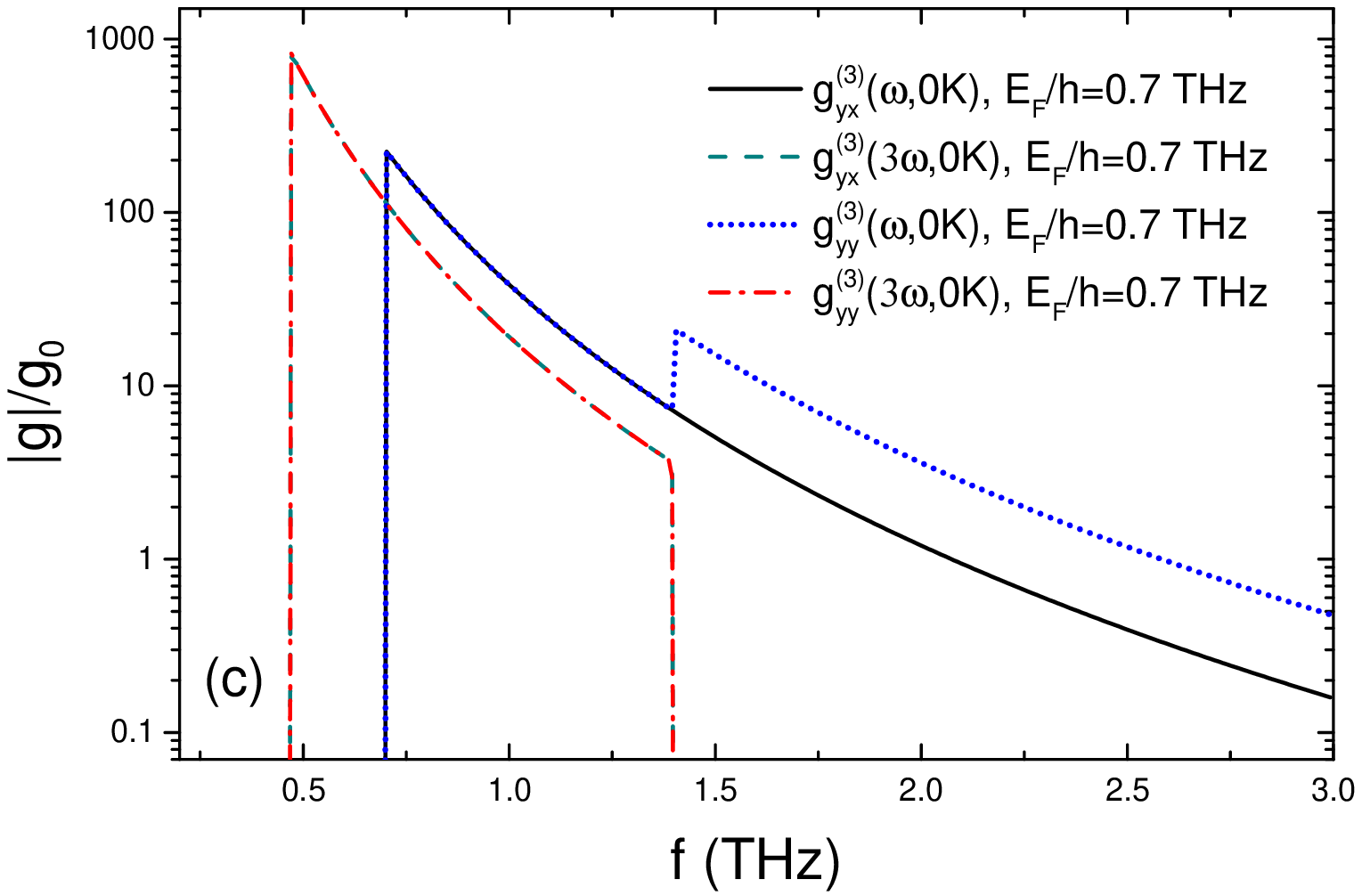}}
\hfil
\subfloat{\label{fig:7d}\includegraphics[width=0.5\linewidth]{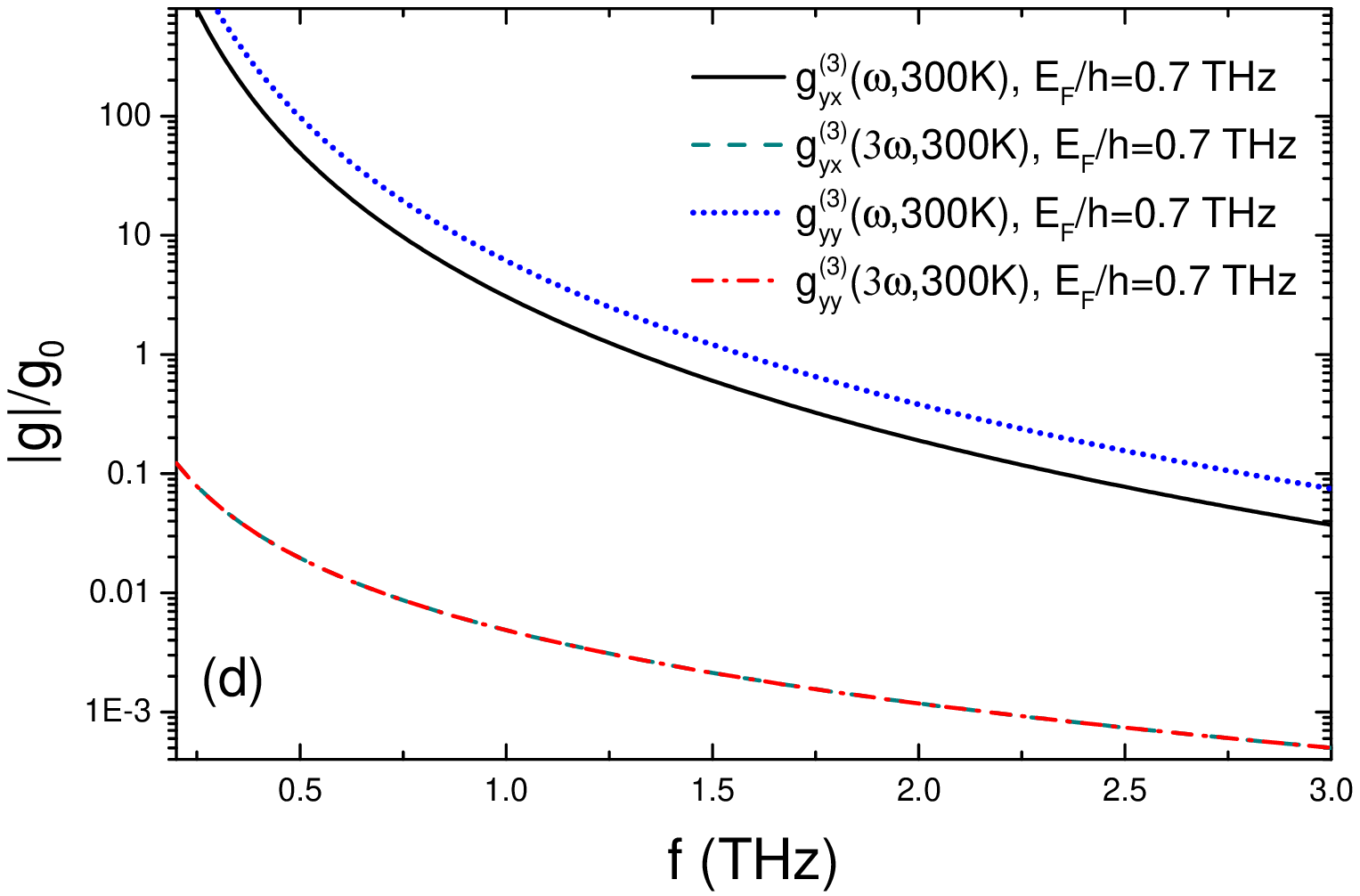}}
\caption{(color online) Comparison of isotropic Kerr and third-harmonic nonlinearities of
extrinsic acGNR20 $(E_F /h = 0.7 \, \mathrm{THz})$ at (a)
$T = 0 \, \mathrm{K}$; and (b) $T = 300 \, \mathrm{K}$ with those of intrinsic 2D SLG; and
comparison of isotropic and anisotropic Kerr and third-harmonic nonlinearities
of extrinsic acGNR20 $(E_F /h = 0.7 \, \mathrm{THz})$
at (c) $T = 0 \, \mathrm{K}$; and (d) $T = 300 \, \mathrm{K}$. The field
strength used in all calculations is $E_y = 10 \, \mathrm{kV/m}$ and the
excitation frequency $f = \omega/2 \pi$.}
\label{fig:7}
\end{figure}

Figs. \ref{fig:2}-\ref{fig:7} present results computed using our model described
in Section \ref{Model}. Fig. \ref{fig:2} summarizes the comparison between the
results for intrinsic 2D SLG and acGNR, indicating that at low temperatures, the isotropic
third-order Kerr conductances is significantly larger than for 2D SLG. At $T = 0
\, \mathrm{K}$, the third-order third-harmonic conductance is zero. The room
temperature Kerr conductance continues to be signficantly larger, and the
third-harmonic conductance becomes of the order of that for 2D SLG.
Fig. \ref{fig:3} describes both the temperature and width dependence of the
third-order conductances for thin, metallic acGNR. The decay with increasing temperature for
the acGNR Kerr conductances are similar to that of 2D SLG, with the acGNR
conductances maintaining their significantly larger relative size. For the
third-harmonic conductances, quite different behavior is observed; the acGNR
third-harmonic conductance is $0$ at $T=0\,\mathrm{K}$, increases to a maximum,
and then decays much faster than for 2D SLG with further increases in
temperature. The decay rate as a function of width for all acGNR third-order
conductances is observed to follow a simple width dependence rule discussed
below.

Fig. \ref{fig:4} describes the temperature dependence of the field
strength required for the nonlinear conductance to dominate over the linear
conductance. Results indicate that this critical field is quite small, varying
from $1 - 5 \, \mathrm{kV/m}$ for the third-order Kerr conductance, and
exhibiting a minimum  of $\sim 5 \, \mathrm{kV/m}$ for the third-order
third-harmonic conductance. Figs. \ref{fig:5} and \ref{fig:6} illustrate several
novel features of the Kerr and third-harmonic conductances for
\textit{extrinsic} acGNR as a function of temperature. For the Kerr conductance,
an antiresonance develops at low temperature and broadens with increasing $E_F$.
For the third-harmonic nonlinearity, the antiresonance found at $T = 0 \,
\mathrm{K}$ for intrinsic acGNR is seen to shift to higher temperatures as $E_F$
increases.

Finally, Fig. \ref{fig:7}, illustrates the behavior of the third-order Kerr and
third-harmonic nonlinearities for extrinsic acGNR as a function of excitation
frequency $\omega = 2 \pi f$. Most remarkably, the third-harmonic nonlinearity
is non-zero over a finite bandwidth at $T = 0\, \mathrm{K}$; a result of the
state-blocking that occurs in extrinsic material. The excitation-frequency
dependence of the nonlinear conductances at room temperature is also show. In
the discussion that follows, we investigate each of these features in more detail.

The
frequency dependent nonlinear conductance in units of $g_0=e^2/4\hbar$ for
intrinsic acGNR20, calculated assuming an applied field strength of
\SI{10}{kV/m},
is plotted in Fig. \ref{fig:2}, together with the third-order Kerr conductance
of 2D SLG. Both
nonlinear terms for intrinsic metallic acGNR20 and 2D SLG decrease rapidly with
frequency.  The huge nonlinearities at lower frequencies are associated with the
strong interaction of carriers with low energy photons. The third-order Kerr conductance, $g_{y \nu}^{(3)}(\omega)$
for acGNR20
is approximately three orders of magnitude larger than that for 2D SLG.
The exact enhancement factor for nonlinear conductances in metallic acGNR is a
function of the nanoribbon width, and from Eqs. \ref{eq:gACy}, \ref{eq:gACx}  is determined to be $v_F/
\omega Lx$.
Due to the thermal factor cancellation in the expression for the
nonlinear third-harmonic conductance, $g_{y\nu}^{(3)}(3\omega)$ tends to be much
less than $g_{y\nu}^{(3)}(\omega)$. When $T=\SI{0}{K}$, the third-harmonic
conductance is zero for intrinsic acGNR20. For $T = \SI{300}{K}$, the third-harmonic conductance is of the
same order as for 2D SLG.\par

In Fig. \ref{fig:3}, we illustrate the temperature and width dependence of the
third-order nonlinear conductance for intrinsic metallic acGNR and 2D SLG for an
excitation frequency of \SI{1}{THz} and an applied field strength of \SI{10}{kV/m}.
In Figs. \ref{fig:3a} and \ref{fig:3b}, $g_{y\nu}^{(3)}(\omega)$ is shown to decrease
monotonically with temperature $T$. However, $g_{y\nu}^{(3)}(3\omega)$ is
initially zero at $T = $ \SI{0}{K} and increases to its maximum value ($\sim$ 2 orders
of magnitude above that for 2D SLG) at approximately
$T = $ \SI{17}{K} (the exact location of the maximum is a function of the thermal factor
appearing in the expressions for the conductance). It then decreases at a faster
rate then $g_{y\nu}^{(3)}(\omega)$ for $T>\SI{17}{K}$. The rate of decrease with
temperature for $g_{y\nu}^{(3)}(\omega)$ is approximately the same as for 2D SLG.

In Figs. \ref{fig:3c} and \ref{fig:3d} we see that both third-order nonlinear
conductance components are inversely proportional to the width of the acGNR $L_x$.
This dependence of the conductance on $L_x$ is due to the unitless factor
$v_F/\omega L_x$ in Eqs. \ref{eq:gACy}, \ref{eq:gACx}, which implies that the
total quasi-1D nonlinear current
is constant and invariant of the nanoribbon width. We see that for $L_x \simeq \SI{20}{nm}$, or
acGNR164, $g_{y\nu}^{(3)}(\omega)$ is still greater than that of 2D SLG
for an excitation frequency of
\SI{1}{THz}, which again suggests that thin metallic acGNR ($L_x\leqslant\SI{20}{nm}$) manifests a much
stronger Kerr conductance $g_{y\nu}^{(3)}(\omega)$ than 2D SLG over a wide range of widths. These findings
suggest that metallic acGNR of submicron width is a better candidate than 2D
SLG for nonlinear THz device applications. 

In order to evaluate the frequency-conversion device potential of metallic
acGNR, we define a critical field strength $E_{c,y \nu}^{(3)}(\omega,T)$ as the
field strength when the nonlinear conductance dominates over the linear
conductance ($|g_{y \nu}^{(3)}|/ g_0 >1$ where $g_0 = e^2/4
\hbar$). In Fig. \ref{fig:4}
we plot the temperature dependence of the critical field strength for intrinsic
metallic acGNR assuming a \SI{1}{THz} excitation frequency.
Fig. \ref{fig:4a} illustrates the change in critical field as a function of
temperature for both intrinsic metallic acGNR and 2D SLG. Due to the thermal
factor cancellation, at low temperatures, the third-order conductance $g_{y\nu}^{(3)}(3\omega)$ for acGNR20
exhibits a larger critical field strength than 2D SLG. As the thermal
distribution broadens with increasing $T$, the critical strength drops to $10\%$
of the critical field strength for 2D SLG. As the temperature rises further,
$E_{c,y \nu}^{(3)}(3\omega,T)$ increases until it rises above that for 2D SLG
near $T=$ \SI{170}{K} again. For the Kerr conductance term, the critical field
$E_{c,y \nu}^{(3)}(\omega,T)$, increases as temperature increases, but it stays
$\sim$ 1 order of magnitude below the critical field for 2D SLG. Further, the
relatively small change in critical field for $g_{y\nu}^{(3)}(\omega)$ from $T=$
\SI{0}{K} to $T=$ \SI{300}{K} indicates that metallic acGNR should exhibit
excellent frequency conversion efficiencies for the optical Kerr process. The critical field strength we obtained is much smaller than the damage threshold \cite{bao12}, the strong nonlinear response, or the small values of the critical field exhibited by metallic acGNR for both Kerr and third-harmonic nonlinearities suggest that, low \si{THz} and low doped metallic acGNR are preferable to exploit the nonlinearity at intensities below the damage threshold \cite{nasari16}. As a result, low dopend thin metallic acGNR will be excellent for use in the fabrication of nonlinear optical frequency-conversion devices \cite{coherent10,bao12}.

In Figs. \ref{fig:5} and \ref{fig:6} we study the Kerr
$g_{y\nu}^{(3)}(\omega)$ and third-harmonic $g_{y\nu}^{(3)}(3\omega)$
conductances as a function of the
Fermi level $E_F$ (since the behavior of the system is symmetric for $E_F$ about
$E_F = 0$ in Figs. \ref{fig:5a} and \ref{fig:6a}, we only plot results for positive $E_F$). For $E_F$ well below the optical phonon energy
($\sim$ \SI{200}{meV}), we plot the Fermi-level dependence of $g_{y\nu}^{(3)}(\omega)$ and
$g_{y\nu}^{(3)}(3\omega)$ assuming a \SI{1}{THz} excitation at $T=$ \SI{0}{K}
and $T=$ \SI{300}{K}. Perhaps the most important observations are for the
\SI{0}{K} case. We see three threshold frequencies for $E_F/h$: \SI{0.5}{THz}, \SI{1}{THz} and \SI{1.5}{THz}. These frequencies correspond to turning on/off the
thermal distribution\cite{mik16,cheng1,*cheng1R,cheng2,cheng3,*cheng3R} at $\omega/2$, $\omega$ and $3\omega/2$. We note that
$g_{y\nu}^{(3)}(3\omega)$ is nonzero over the $\omega/2$ to $3\omega/2$ doping
window. In this window, only the $N(\omega)$ thermal factor term
contributes to the $g_{y\nu}^{(3)}(3\omega)$ transition. Near room temperature,
there are always electron and hole states
\cite{cheng1,*cheng1R,cheng2,cheng3,*cheng3R} in the energy range determined by
the thermal factor. As a result, we always observe nonzero conductance at all non-zero
temperatures. This result suggests
that at low temperatures, light doping will greatly enhance
$g_{y\nu}^{(3)}(3\omega)$. But the enhancement we observe at low temperature for
$g_{y\nu}^{(3)}(3\omega)$ disappears near room temperature. Also, the curves for
different values of $E_F$ asymptotically approach the intrinsic acGNR
conductance, as the temperature increases. \par

In Fig. \ref{fig:7}, we compare the conductances $g_{y \nu}^{(3)}(\omega)$ and
$g_{y \nu}^{(3)}(3 \omega)$ of extrinsic acGNR20 ($E_F/h= \SI{0.7}{THz}$) for
different temperatures and with the corresponding values for intrinsic 2D SLG.
For the $T = \SI{0}{K}$ case, we observe a sharp onset for both the isotropic
and anisotropic Kerr conductances at $E_F/h$ ($\omega/2 \pi =\SI{0.7}{THz}$) and
 a further increase at $2E_F/h$ ($\omega/2 \pi =\SI{1.4}{THz}$) for the isotropic Kerr conductance. 
These changes are due to different terms in the thermal factor turning on at these
 excitation frequencies (see Table I).

The third-harmonic result is significantly different at $T=\SI{0}{K}$. In
this case the conductance turns on abruptly at $2E_F/3h$ ($\omega/2 \pi =\SI{0.467}{THz}$)
and turns off abruptly at $2E_F/h$ ($\omega/2 \pi = \SI{1.4}{THz}$). These changes are also due to the relevant terms in
the thermal factor turning on at particular excitation frequencies (see Table
I).

\begin{table}[!htbp]
\centering
\caption{Thermal Factor Terms for excitation frequency $\omega$ (\textit{cf.}
Eqs. \ref{eq:gACy}, \ref{eq:gACx})}
\label{table:i}
\begin{tabular}{cc}
\hline\hline
\multicolumn{2}{c}{Kerr Conductance $(T = 0 \, \mathrm{K})$} \\
\hline\hline
Frequency Range & Thermal Factor Terms \\
\hline
$0 < \omega \leq E_F/\hbar$ & all terms are 0 \\
$E_F/\hbar < \omega \leq 2 E_F/\hbar$ & $N(\omega,E_F) =1 $ \\
$\omega > 2 E_F/\hbar \text{, isotropic}$ & $2N(\omega/2,E_F) + N(\omega,E_F) =
3$ \\
$\omega > 2 E_F/\hbar \text{, anisotropic}$ & $ N(\omega,E_F) = 1$\\
\hline\hline
\multicolumn{2}{c}{Third-harmonic Conductance $(T = 0 \, \mathrm{K})$} \\
\hline\hline
Frequency Range & Thermal Factor Terms \\
\hline
$0 < \omega \leq 2 E_F/3\hbar $ & all terms are 0 \\
$2 E_F/3\hbar  < \omega \leq E_F/\hbar$ & $-\frac{1}{2}N(3 \omega/2,E_F) =
-\frac{1}{2}$ \\
$E_F/\hbar < \omega \leq 2 E_F/\hbar$ & $ N(\omega,E_F) -\frac{1}{2} N(3
\omega/2,E_F) = \frac{1}{2}$\\
$\omega > 2 E_F/\hbar$ & $-\frac{1}{2}N(\omega/2,E_F) +  N(\omega,E_F)
-\frac{1}{2} N(3 \omega/2,E_F) = 0$ \\
\hline
\end{tabular}
\end{table}

For $T = 300 \, \mathrm{K}$, we note that the extrinsic Kerr conductance is strongly
enhanced over intrinsic 2D SLG, as it is in the intrinsic case. Further, the extrinsic
third-harmonic conductance is of the same order as the 2D SLG nonlinear Kerr conductance value.
Comparing the isotropic conductances with their anisotropic counterparts, we
note similar behavior at $T = 300 \, \mathrm{K}$.
These results indicate that for low temperatures, there is a strong enhancement
of the third-harmonic nonlinearity; however at room temperature, the Kerr
nonlinearity dominates. 

Finally, it is worth noting the limitations of our approach. The singularity around the
Dirac point in metallic acGNR leads to high mobility, but acGNR can be more prone
to edge defects. Furthermore the $\mathbf{k} \cdot \mathbf{p}$ approximation is
appropriate only at low energies, well below
\SI{2}{eV}.\cite{raza12}  For Fermi energies greater than optical phonon energy
\SI{200}{meV}, one needs to use a more basic tight-binding description, and the Dirac physics becomes largely
irrelevant\cite{raza12}. For undoped and lightly-doped acGNR, the Fermi energy is well away from these
energy scales and the description in terms of the Dirac Hamiltonian should work
relatively well.  In this paper, 
we assume there is no coupling of the local nonlinear current density with the
spatial distribution of the applied electric field.
Further, we treat the metallic acGNR with no applied longitudinal bias
voltage, so that
the Fermi level does not change across the longitudinal direction of the
nanoribbon. It will
be important to introduce additional effects in the present model such as
the finite extent of the excitation field and the finite longitudinal size of the
nanoribbon, as well as material effects such as electron-electron,
electron-phonon interactions, and other edge
effects. These topics are the subject of our future work.

\section{selection rules related to acGNR}
In this section, we discuss the applicability of well-known selection rules for
acGNR and 2D SLG to the problem of THz nonlinear harmonic generation in thin metallic
acGNR. We focus on the interband transition in the lowest (linear) band ($n=0$).
The fact that we have nonzero $g_{yy}$ and zero $g_{xx}$ is consistent with the
selection rules for acGNR found by
Sasaki \textit{et.al.}\cite{gnrselectionrule} and HC Chung
\textit{et.al.}\cite{gnrselectionrule2}\par

In general, for 2D SLG there is no anisotropic current ($J_y, \, J_x$ induced by
$E_x, \,E_y$).
The anisotropic conductance for intrinsic 2D SLG can be written:
\begin{equation}
\begin{aligned}
g_{yx}^{(1)}(\omega)_{2D} &=\lim_{\Gamma\to0}-\frac{g_0}{ \pi^2} \int_{0}^{2 \pi}
\sin(2\theta) d \theta \int_{0}^{\infty}\Re \left[ i \frac{v_F^2}{\omega^2}
\frac{k v_F}{2 k v_F-\omega-i \Gamma} k \tanh(\frac{\hbar v_F k}{2 k_B T}) \right]dk \\
g_{yx}^{(3)}(\omega)_{2D} &=\lim_{\Gamma\to0}\frac{g_0}{ \pi^2} \frac{\eta^2}{2}
\int_{0}^{2 \pi}  \sin(2\theta) d \theta \\
&\times \int_{0}^{\infty}  \Re \left[ i\frac{v_F^2}{\omega^2}
\frac{k^2 v_F^2 [- k v_F+\omega+k v_F \cos(2 \theta)]}{[(2 k v_F-\omega)^2+\Gamma^2](k v_F-\omega-i\Gamma)}
k \tanh(\frac{\hbar v_F k}{2 k_B T}) \right] dk \\
g_{yx}^{(3)}(3\omega)_{2D} &=\lim_{\Gamma\to0}\frac{g_0}{ \pi^2} \frac{\eta^2}{6} \int_{0}^{2
\pi}  \sin(2\theta) d \theta \\
&\times \int_{0}^{\infty}  \Re \left[ i\frac{v_F^2}{\omega^2}
\frac{k v_F [k^2 v_F^2-3 k v_F \omega+4 \omega^2-k^2 v_F^2
\cos(2 \theta)]}{(2 k v_F-\omega-i\Gamma)(k v_F-\omega-i\Gamma)(2 k v_F-3\omega-i3\Gamma)} k
\tanh(\frac{\hbar v_F k}{2 k_B T}) \right] dk 
\end{aligned}
\end{equation}
where $\eta=\frac{e  A_y v_F}{\hbar \omega}=\frac{e E_y v_F}{\hbar \omega^2}$ measures the e-h coupling strengh.
Using this result, we see that because $ \int_{0}^{2 \pi}  \sin(2\theta)=0$, the
conductance terms $g_{yx}^{(i)}(\omega_0)_{2D}=0$ for 2D SLG. The
$g_{xy}^{(i)}(\omega_0)_{2D}=0$ from similar analysis. The zero anisotropic
current in 2D SLG results from that fact that the net sum is zero over all
possible angles, and agrees with the quantum analysis performed in Ref. \cite{mik16} for 2D SLG.\par

However, as shown above for metallic acGNR, $J_{\nu}$, $\sigma_{\mu
\nu}^{(i)}(m,\omega)$ has the general form:
\begin{equation}
\sigma_{y\nu}^{(i)}(m,\omega) =F_{y\nu}^{(i)}(|k_y|) \cos(\theta_{k_n,k_y})
\end{equation}
For metallic acGNR, we no longer integrate all possible angles as we did for 2D SLG.
Due to the 1D nature of acGNR, we only have $\theta_{k_n,k_y}=0,\pi$ depending
on the sign of $k_y$, and thus we only evaluate at two angles according to the
initial condition given by Eq. \ref{i.c.} when we evaluate the total current
density $J_{\nu}$ for metallic acGNR. As a result, $J_{\nu}$ is not always
zero for all $E_F$, $\omega$ and $T$. For direct interband
transitions between states where $k_{x,n}\neq 0$, we make a similar argument as
we only require states at $\epsilon=\hbar v_F k_{x,n} \csc(\theta_{k_n,k_y})$ to be at resonance.
Thus, we only have the $\theta_{k_n,k_y}$ and $\pi-\theta_{k_n,k_y}$ pair as the two
solutions. In this way, we extend the selection rules of the direct interband
transition for acGNR to the $J_{x}$ case, \textit{i.e.} $k_{x,n}$ does not change
from initial state to final state. This is the same requirement as for $J_{y}$
in acGNR.

\section{Concluding Remarks}\label{Conclusions}
Kimouche \textit{et.al.}\cite{kimouche2015ultra} and Jacobberger \textit{et.al.}
\cite{jacobberger2015direct} have succesfully fabricated ultrathin, smooth acGNR with
widths $L_x < 10 \, \mathrm{nm}$. Our calculation of the
nonlinear conductance in acGNR suggest that experimental measurements of the
THz nonlinear response in thin
metallic acGNR should be measurable at relatively low excitation field strengths.
The relatively small critical field
strength at room temperature implies that thin metallic acGNR have significant potential
for nonlinear device applications. The striking turn on and turn off of the
third-order harmonics with small changes in Fermi level at low temperatures
suggest that metallic acGNR could be the the basis for developing a sensitive
graphene-based low temperature detector or oscillator.\par

In this paper, we have modeled the third-order THz response of metallic acGNR
using a nonlinear semi-analytical approach. The time-dependent Dirac equation
for massless Dirac Fermions is solved via the Fourier expansion method.  We have shown
intrinsic metallic acGNR exhibits  strong nonlinear effects from the THz to the FIR
regime under applied electric field amplitudes less than \SI{10}{kV/m}.
We also describe the behavior of these nonlinearities for extrinsic, metallic
acGNR. Under certain
conditions, metallic acGNR will exhibit a larger nonlinear conductance, require
less applied electric field strength to generate moderate strong high harmonics
and show better temperature stability than intrinsic 2D SLG. This opens the
potential for use in many device applications for intrinsic and slightly doping metallic acGNR.

\appendix
\section{Vector Potential}\label{app1}
In the Coulomb gauge, for a constant scalar potential $(\nabla \varphi=0)$, the
relationship between the vector potential and the electric field is
$E(t) = - \partial A(t)/\partial t$. Thus, for an electric field that is turned on at
time $t_0$, the vector potential is written:
\begin{equation}\label{eq:Adef}
A(t) = - \int_{t_0}^t E(t_1) \, dt_1 = -E_0 \int_{t_0}^t e^{-i \omega t_1} \, dt_1
\end{equation}
Considering a time-harmonic field turned on
at $t_0 \to -\infty$, we write the integral in \cref{eq:Adef}:
\begin{equation}
\begin{aligned}\label{eq:A2}
I=&\int_{-\infty}^{t} e^{-i \omega t_1}\, d t_1 \\
=&\int_{-\infty}^{0} e^{-i \omega t_1}\, d t_1+\int_{0}^{t} e^{-i \omega t_1}\,
d t_1 \\
=&I_1+I_2
\end{aligned}
\end{equation}
In order to evaluate the integral $I_1$, we introduce an infinitesimally small
positive parameter $\tau$, which corresponds to the field turning on
adiabatically \cite{ryu07,mahan} at $-\infty$. With $t'=-t_1$:
\begin{equation}
\begin{aligned}
I_1 &= \lim_{\tau \to 0} \int_{-\infty}^0 e^{(\tau - i \omega) t_1} \, dt_1 \\
&=\lim_{\tau\to0}\int_{0}^{\infty} e^{-(\tau-i\omega)t'}\, d t'\\
&= \lim_{\tau \to 0} \frac{1}{\tau - i \omega} = \frac{1}{-i \omega}
\end{aligned}
\end{equation}

Evaluating the integral $I_2$, we obtain:
\begin{equation}
\begin{aligned}
I_2 &= \int_0^t e^{-i \omega t} \, dt_1 \\
&= \frac{e^{-i \omega t}-1}{-i \omega}
\end{aligned}
\end{equation}
The total integral $I$ is obtained by summing $I_1$ and $I_2$:
\begin{equation}
I = I_1 + I_2 = \frac{1}{-i \omega} + \frac{e^{-i \omega t} -1}{-i \omega} =
\frac{e^{-i \omega t}}{-i \omega}
\end{equation}
and the vector potential in the Coulomb gauge for a time-harmonic electric field
that turns on adiabatically at $t_0 \to -\infty$ becomes:
\begin{equation}
A(t) = - E_0 I = \frac{-E_0 e^{-i \omega t}}{-i \omega} = \frac{E(t)}{i \omega}
\end{equation}

\section{Derivation of the 2D SLG Nonlinear Conductance}
Following Wright, \textit{et.al.}\cite{wright09} and Ang \textit{et.al.}
\cite{ang2015nonlinear}, we compute the third-order current
densities for 2D SLG
due to an $\hat{x}$-polarized electric field of the form $\hat{x} E_0 e^{i
\omega t}$. Defining $p=\sqrt{p_x^2+p_y^2}$, and $\tan(\theta)=\frac{p_y}{p_x}$,
and using the fact that $\int_0^{2\pi} \cos(2\theta) d\theta=\int_0^{2\pi} \sin(2\theta) d\theta=\int_0^{2\pi}\cos(4\theta) d\theta=\int_0^{2\pi}\sin(4\theta) d\theta=0$
the current densities in the $\hat{x}$ direction are written\cite{wright09,ang2015nonlinear,lopez08,blg10,subgap11,lopez08,mishchenko09,bao12,zhou11}:
\begin{align}
J_3^x(\omega)=&\lim_{\Gamma\to0}\frac{g_sg_v}{(2\pi\hbar)^2}g_0 \eta
\int_{0}^{2\pi}d\theta \int_0^{\infty} \Re\left\{i\frac{-v_F^2 (3p^3v_F^3-8\hbar
p^2v_F^2\hbar\omega+6p v_F
\hbar^2\omega^2-2\hbar^3\omega^3)N(p)p}{\omega^2[2pv_F-\hbar(\omega+i\Gamma)][2pv_F-\hbar(\omega-i\Gamma)][pv_F-\hbar(\omega+i\Gamma)]}\right\}dp
\label{eq:j3x2dslgw}\\
J_3^x(3\omega)=&\lim_{\Gamma\to0}\frac{g_sg_v}{(2\pi\hbar)^2}g_0 \eta
\int_{0}^{2\pi}d\theta \int_0^{\infty} \Re\left\{i\frac{v_F^2 (3p^3v_F^3-12\hbar
p^2v_F^2\hbar\omega+14p v_F
\hbar^2\omega^2-6\hbar^3\omega^3)N(p)p}{3\omega^2[2pv_F-\hbar(\omega+i\Gamma)][pv_F-\hbar(\omega+i\Gamma)][2pv_F-3\hbar(\omega+i\Gamma)]}\right\}dp\label{eq:j3x2dslg3w}
\end{align}
with $g_s, \, g_v=2$, $g_0=\frac{e^2}{4\hbar}$, $\eta=\frac{e^2
E_0^2 v_F^2}{\hbar^2\omega^4}$, and $N(p)=\tanh(\frac{p v_F}{2k_BT})$.\par

In these expressions, the integrands are of the form:
\begin{align}
i_1(x) &= f_1(x) \Re \left[ i \frac{1}{(2x-x_0-i \Gamma)(2x-x_0+i \Gamma)(x-x_0-i \Gamma)}
\right] \\
i_3(x) &= f_3(x) \Re \left[ i \frac{1}{(2x-x_0-i \Gamma)(x-x_0-i \Gamma)(2x-3x_0-i3 \Gamma)}
\right]
\end{align}
for the Kerr and third-order currents respectively,
with $f_1(x)$, $f_3(x)$, $x_0$, $\Gamma$ real.
After some algebra we find that these integrands become:
\begin{subequations}
\begin{align}
i_1(x)  =& f_1(x) \frac{\pi}{x\left(3x-2x_0\right)}
\left[ \frac{1}{\pi} \frac{\Gamma}{(2x-x_0)^2 + \Gamma^2} - \frac{1}{\pi}
\frac{\Gamma}{(x-x_0)^2+\Gamma^2}
\right]\\
i_3(x)=& f_3(x) \frac{\pi}{x^2}
\left[ -\frac{1}{4} \frac{1}{\pi} \frac{\Gamma}{(2x-x_0)^2 +
\Gamma^2}  + \frac{1}{\pi}
\frac{\Gamma}{(x-x_0)^2 + \Gamma^2} -\frac{9}{4}
\frac{1}{\pi} \frac{3\Gamma}{(2x-3x_0)^2 + 9\Gamma^2} \right]
\end{align}\label{eq:integrand}
\end{subequations}
As a result, the expressions for the current density in
\cref{eq:j3x2dslgw,eq:j3x2dslg3w} above may
be expanded as a set of integrals of the form:
\begin{equation}
Z_1=\lim_{\Gamma\to0}\int_a^b \Re\left[\frac{iz(x)}{x-x_0\mp i \Gamma}\right]\,dx=\lim_{\Gamma\to0}\int_a^b \Re\left[z_1(x,x_0,\Gamma)\right]\,dx
\end{equation}
with $z(x)$, $x$, $x_0$, $\Gamma>0$ real. Using the property:
\begin{equation}\label{eq:simpleresult}
\lim_{\Gamma \to 0}\frac{1}{\pi}\frac{\Gamma}{(x-x_0)^2+\Gamma^2}=\delta(x-x_0)
\end{equation}
we arrive at $Z_1=\pi f(x_0)$. Several example problems involving this type of
kernel may be found in Refs. \cite{delta1,delta2,delta3,delta4}.

Alternatively, we may use the Cauchy Principal Value theorem to solve this
problem. Separating the real and imaginary parts of the integrand
$z_1(x,x_0,\Gamma)$, we obtain:
\begin{align}
\Re\left[z_1(x,x_0,\Gamma)\right]&=\mp\pi \frac{1}{\pi}\frac{\Gamma}{(x-x_0)^2+\Gamma^2} z(x)\\
\Im\left[z_1(x,x_0,\Gamma)\right]&=\frac{1}{(x-x_0)+\Gamma^2/\left(x-x_0\right)}z(x)
\end{align}
The Sokhotsky-Plemelj theorem on the real interval $[a,b]$
states\cite{sokhotsky}:
\begin{equation}
\lim_{\Gamma\to0}\int_a^b \frac{g(x)}{x-x_0\mp i \Gamma}dx=\mathbb{P} \int_a^b \frac{g(x)}{x-x_0}dx \pm i\pi g(x_0)
\end{equation}
where $\mathbb{P} \int_a^b g(x) dx$ denotes the Cauchy principal integral of $g(x)$.
For $g(x)=iz(x)$ with $z(x)$ real, the real and imaginary parts become:
\begin{align}
\Re\left[\lim_{\Gamma\to0}\int_a^b \frac{iz(x)}{x-x_0\mp i \Gamma}\,dx\right]&=\mp \pi z(x_0)\\
\Im\left[\lim_{\Gamma\to0}\int_a^b \frac{iz(x)}{x-x_0\mp i \Gamma}\,dx\right]&= \mathbb{P} \int_a^b \frac{z(x)}{x-x_0}dx
\end{align}
which is the same result as in \cref{eq:simpleresult}.

An analysis of the interband transition using the Kubo formula has appeared in
Ref. \cite{eo10}. Eq. (A1) of that reference further confirms our
result for 2D SLG.

Based on the above analysis, in the limit as $\Gamma \to 0$, the integrands in \cref{eq:integrand} reduce to:
\begin{align}
\lim_{\Gamma\to0} i_1(x) &= \frac{\pi f_1(x)}{x(3x-2x_0)} \left[\frac{\delta(x-\frac{x_0}{2})}{2}
 - \delta(x-x_0) \right] \nonumber \\
&=-\frac{\pi f_1(x)}{x_0^2} \left[ 2\delta(x-\frac{x_0}{2})+\delta(x-x_0) \right]\\
\lim_{\Gamma\to0} i_3(x) &=\frac{\pi f_3(x)}{x^2}\left[-\frac{1}{4}\frac{\delta(x-\frac{x_0}{2})}{2}+\delta(x-x_0)-\frac{9}{4}\frac{\delta(x-\frac{3x_0}{2})}{2} \right]
\end{align}
and the integrals reduce to:
\begin{align}
& I_1 = -\frac{\pi}{x_0^2}[2f_1(\frac{x_0}{2})+f_1(x_0)] \\
& I_3
=-\frac{\pi}{2x_0^2}\left[f_3(\frac{x_0}{2})-2 f_3(x_0)+f_3(\frac{3x_0}{2})\right]
\end{align}
Therefore, the current densities may be written:
\begin{align}
 J_3^{x}(\omega)&=-g_0 E_0 \frac{e^2 E_0^2 v_F^2}{\hbar ^2 \omega ^4} \left[
\frac{5}{4}\tanh(\frac{\hbar \omega}{4 k_B T})+2\tanh(\frac{\hbar \omega}{2
k_B T}) \right]\\
 J_3^{x}(3\omega)&=g_0 E_0 \frac{e^2 E_0^2 v_F^2}{\hbar ^2 \omega ^4} \left[
\frac{13}{48}\tanh(\frac{\hbar \omega}{4 k_B
T})-\frac{2}{3}\tanh(\frac{\hbar \omega}{2 k_B T})+
\frac{45}{48}\tanh(\frac{3\hbar \omega}{4 k_B T})\right]
\end{align}
resulting in the Kerr conductance:
\begin{equation}\label{eq:kerrconductance}
g_{xx}^{(3)}(\omega)_{2D}=-g_0 \frac{e^2 E_0^2 v_F^2}{\hbar ^2 \omega ^4} \left[
\frac{5}{4}\tanh(\frac{\hbar \omega}{4 k_B T})+2\tanh(\frac{\hbar \omega}{2
k_B T}) \right]
\end{equation}
and the third-harmonic conductance:
\begin{equation}
g_{xx}^{(3)}(3\omega)_{2D}=g_0 \frac{e^2 E_0^2 v_F^2}{\hbar ^2 \omega ^4} \left[
\frac{13}{48}\tanh(\frac{\hbar \omega}{4 k_B
T})-\frac{2}{3}\tanh(\frac{\hbar \omega}{2 k_B T})+
\frac{45}{48}\tanh(\frac{3\hbar \omega}{4 k_B T})\right]
\end{equation}
Similarly, for a $\hat{y}$-polarized electric field of the form $\hat{y}
E_0
e^{i \omega t}$, we arrive at an identical result for the third-order Kerr
current in the $\hat{y}$ direction, or equivalently $g_{xx}^{(3)}(\omega)
=g_{yy}^{(3)}(\omega)$ and $g_{xx}^{(3)}(3 \omega) =g_{yy}^{(3)}(3 \omega)$ for 2D SLG.

\begin{figure}[H]
\centering
\includegraphics[width=1.0\textwidth]{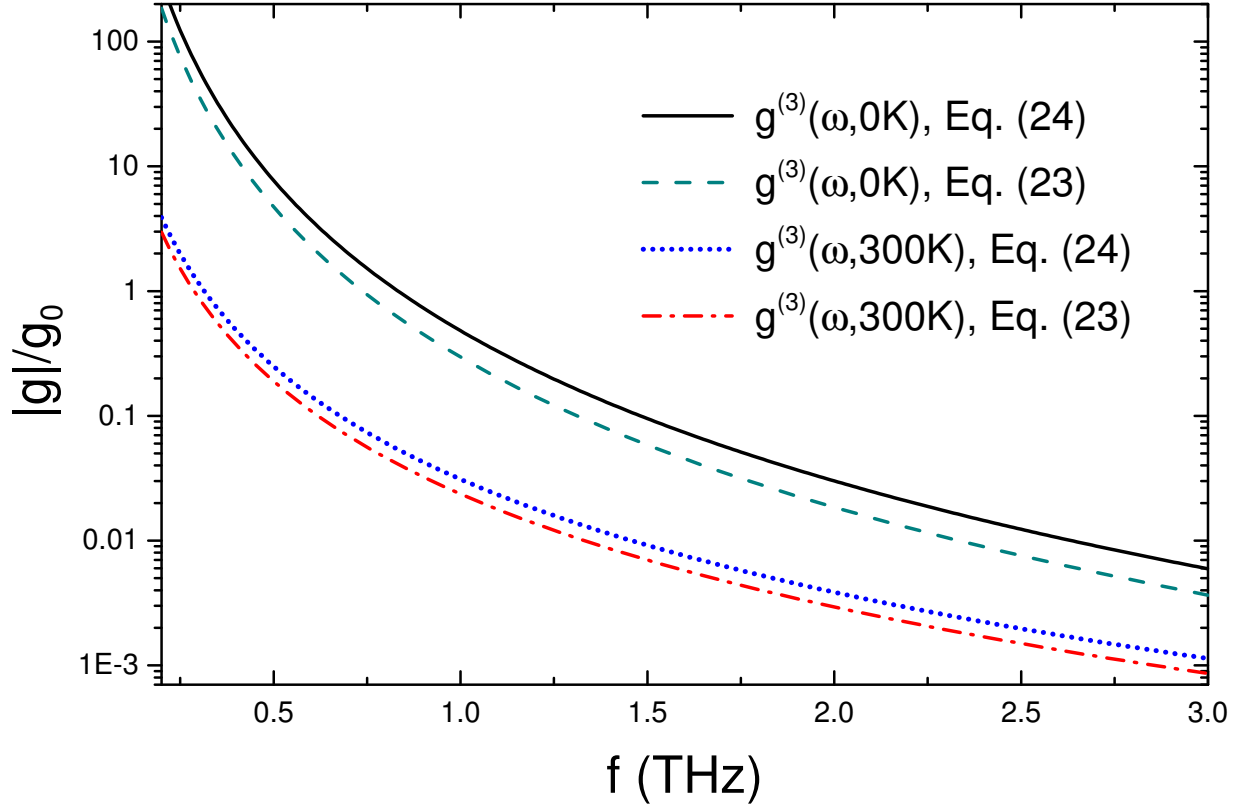}
\caption{(color online) Comparison of the third-order nonlinear conductance of
for intrinsic 2D SLG from \cref{2Dg3w1v1,2Dg3w1v2} at $T = \SI{0}{K}$ and \SI{300}{K}.
The field strength used in all calculations is $E_0 = \SI{10}{kV/m}$ and the
excitation frequency $f = \omega/(2 \pi)$.}
\label{fig:8}
\end{figure}

A comparison of \cref{2Dg3w1v2,2Dg3w1v1} for intrinsic 2D SLG with
$E_0=\SI{10}{kV/m}$ at $T = \SI{0}{K}$ and \SI{300}{K} is plotted in \cref{fig:8}.
This shows that while small, the correction due to the $\omega/2$ resonant term
is certainly not negligible.

To further amplify our point that there are two terms in the expression for the
third-order Kerr
nonlinear conductance, we note that Eq. 33 may also be written as:
\begin{equation}\label{eq:symmetricintegrand}
\begin{aligned}
i_1(x) &= f_1(x) \Re \left[ i \frac{1}{(2x-x_0-i \Gamma)(2x-x_0+i
\Gamma)(x-x_0-i \Gamma)} \right] \\
&=- f_1(x) \left[ \frac{1}{(2x-x_0)^2 + \Gamma^2} \right]
\left[ \frac{\Gamma}{(x-x_0)^2 + \Gamma^2} \right] \\
&= -\frac{f_1(x) \Gamma}{4} \left[ \frac{1}{(x-a_1 x_0)^2 + (a_1\Gamma)^2}
\right] \left[
\frac{1}{(x-a_2 x_0)^2 + (a_2\Gamma)^2} \right]
\end{aligned}
\end{equation}
with $a_1=1/2, a_2=1$. Eq. \ref{eq:symmetricintegrand} is symmetric in
$(a_1,a_2)$, and therefore the integral $I_1$ must also be symmetric in
$(a_1,a_2)$. Thus, both $\omega/2$ and $\omega$ terms must appear in the
expression for the Kerr conductance, Eq. \ref{eq:kerrconductance}.

\bibliographystyle{apsrev4-1}
\nocite{*}
%

\newpage
\par
\par
\par
\par
\par

\end{document}